\documentclass[twocolumn,showpacs,preprintnumbers,amsmath,amssymb,aps]{revtex4}

\usepackage{graphicx}              
\usepackage{color}
\usepackage{amssymb}

\begin{document}

\title{The Effect of Mechanical Resonance on Josephson dynamics
}

\author{C. Padurariu}
\author{C. J. H. Keijzers} 
\author{Yu. V. Nazarov}
\affiliation{Kavli Institute of NanoScience, Delft University of Technology, Lorentzweg 1, 2628 CJ, Delft, The Netherlands.}
\date{\today}


\begin{abstract}

We study theoretically dynamics in a Josephson junction coupled to a mechanical resonator looking at the signatures of the resonance in d.c. electrical response of the junction. 
Such a system can be realized experimentally as a suspended ultra-clean carbon nanotube brought in contact with two superconducting leads.
A nearby gate electrode can be used to tune the junction parameters and to excite mechanical motion. 
We augment theoretical estimations with the values of setup parameters measured in the samples fabricated. 

We show that charging effects in the junction give rise to a mechanical force that depends on the superconducting phase difference. The force can excite the resonant mode provided the superconducting current in the junction has oscillating components with a frequency matching the resonant frequency of the mechanical resonator. We develop a model that encompasses the coupling of electrical and mechanical dynamics. We compute the mechanical response (the effect of mechanical motion) in the regime of phase bias and d.c. voltage bias. We thoroughly investigate the regime of combined a.c. and d.c. bias where  Shapiro steps are developed and reveal several distinct regimes characteristic for this effect.
Our results can be immediately applied in the context of experimental detection of the mechanical motion in realistic superconducting nano-mechanical devices.

\end{abstract}

\maketitle


\pagenumbering{arabic}

\section{Introduction}
\label{intro}

Nanoscale electromechanical systems (NEMS) convert small amplitude mechanical motion into measurable electrical currents \cite{Clelandbook}. Devices based on NEMS have found applications as sensitive detectors of mass \cite{Eki2004}, force \cite{Blen2007} and electrical charge \cite{Cle98}. Considerable research efforts have been dedicated to improving detection sensitivity by fabricating devices with higher resonance frequencies, lower damping rates (high quality factors) and larger coupling between electrical and mechanical degrees of freedom. 

The problem of detecting the quantum state of a macroscopic mechanical resonator gave rise to several measuring schemes, proposed \cite{Armour, Schmidt} as well as realized \cite{Cleland}. Continuous improvements in device fabrication in combination with new techniques for cooling mechanical motion \cite{Marq, Son} have pushed the sensitivity threshold to the quantum limit \cite{Cleland}. The use of superconducting devices, in particular, superconducting qubits to detect and control the mechanical motion is in focus  of modern research \cite{Rou2009, Cleland}.  It gives rise to a growing interest in techniques of coupling NEMS to superconducting circuits. 

Superconducting nano-devices frequently use Coulomb blockade that makes their transport properties sensitive to the gate voltages. The same gate voltage can be used to excite the mechanical motion
which is detected from the change of d.c. transport properties of the device \cite{McEuwen, Herre, Gary}. Without superconductors, this scheme has been successfully realized for a metallic single-electron transistor
\cite{Naik} and for a Coulomb-blockaded quantum dot in an ultra-clean carbon nanotube (CNT) \cite{Gary}.
The results revealed high resonance frequencies, reaching gigahertzs, and unprecedented quality factors of the order of $10^5$.
These devices can be made superconducting by connecting them to superconducting leads and providing sufficiently large coupling between the states of the lead and device.
We have successfully realized Josephson junctions based on the ultra-clean CNT. The supercurrent observed demonstrates a pronounced gate-voltage sensitivity that indicates a well-developed Coulomb blockade \cite{Han-unpublished}.

A very interesting proposal that combines Josephson dynamics and mechanical resonator has been recently put forward by Gothenburg collaboration \cite{Sonne}. The authors consider an ideal ballistic CNT between two superconducting leads biased at voltage $V$. Owing to Josephson relation, the current in the nanotube oscillates at frequency $\omega_j = 2eV/\hbar$. The authors notice that in external magnetic field this gives rise to an oscillating Lorentz force. If the frequency matches the frequency of the mechanical resonator, the force excites mechanical motion which rectifies the Josephson current enabling the observation of the effect in d.c. electric response of the junction. One would observe a narrow current peak in $I-V$ characteristics
of the device. The same mechanism is responsible for  Fiske steps \cite{Fiske}: the difference is that in Fiske experiments the resonance is electrical rather than mechanical.

This provides us motivation for the present theoretical study where we address 
superconducting NEMS where a mechanical resonator is integrated with a superconducting circuit element, a Josephson junction. 
The focus of this work is to describe the techniques of driving and detecting mechanical motion using superconducting current and to explore the coupled dynamics of the oscillator displacement and superconducting phase difference. While the study is general for any type of oscillating nanowire, for the purpose of illustration and for concrete estimations we refer to one of the more successful NEMS devices: a suspended metallic CNT resonator connected to the metallic leads. We have fabricated and studied such devices. 

Studies of CNT Josephson junctions have shown that their Josephson energy can be modulated by the gate-induced charge on the CNT. In this paper, we have shown that this gives rise to a different and generally more important mechanism of mechanical driving than that considered in \cite{Sonne}. We consider all possible  non-linearities in coupled Josephson-mechanical dynamics. We have shown that in our situation the most important one is the mechanical non-linearity. We provide detailed  estimations of the displacement, force and electrical current scales involved. It is our conclusion that the mechanical response in our devices should be small modifying the current 
on the scale of $10^{-3}$ of the critical current (In this case, the critical current is estimated from the product of the junction conductance and superconducting energy gap and can exceed the experimentally measured switching current by two orders of magnitude).

In the present paper, we have investigated in detail the mechanical response under conditions of phase bias and d.c. voltage bias. We dedicated special attention to the dynamics in the presence of external a.c. drive, in the regime where Josephson junction exhibits well-developed Shapiro steps \cite{Shapiro}. One of the motivations of the research on Shapiro steps was the better synchronization conditions in comparison with d.c. voltage bias. The reason for this is that the big quality factor $Q$ of the nanomechanical resonance results in a narrow Fiske-like current peak. Its width in voltage can be estimated as $\delta V \simeq V/Q$. 
This imposes a severe limitation on voltage noise $S_V$: to resolve the peak one must achieve
$S_V \ll (e/\hbar) V/Q$. This may be challenging under realistic experimental circumstances. 
There is a way out: the voltage can be synchronized with the frequency of external irradiation. This effectively reduces the noise. From the other hand, the external irradiation can excite the mechanical motion by itself. The resulting complicated dynamics should be augmented with non-linear effects.

In this paper we present our theoretical results concerning the mechanical response manifested in d.c. electric response of the junction, in particular, the modification of  the width and position of Shapiro steps, and analyze a variety of distinct regimes that differ in relative and absolute magnitudes of driving forces and resonant conditions, and are manifested in distinct lineshapes of the resonant response. Our preliminary experimental results show corresponding features. They will be presented elsewhere \cite{Han-unpublished} upon completion of detailed analysis and comparison with our theoretical findings.

The paper is organized as follows. 

In  Section \ref{prop} we describe the setup concentrating separately on the electrical and mechanical properties in the corresponding Subsections and introduce the notations. In Section \ref{Fe} we analyze in detail the mechanisms of coupling between mechanical and electrical degrees of freedom. We consider electrostatic energy, derive and explicate the concept of {\it Josephson mechanical force}. We present a rather involved analysis of competing non-linearities and conclude that for our devices the mechanical non-linearities dominate and the Josephson force can be strong enough to induce the non-linear mechanical response. We give the workflow we use to compute the mechanical response. In the end of the Section, we specify a set of concrete parameters based on experimentally measured values. In Section \ref{sec:phasebias} we address the phase bias conditions. We reveal that the {\it phase-dependent shift} of the resonant frequency can be quite noticeable and discuss Lorentz-like and Fano-like frequency dependences of the mechanical response.
In Section \ref{sec:dcvoltage} we discuss the Fiske-type mechanical response at Josephson frequency matching the resonance frequency of the mechanical resonator, or an integer fraction of this frequency by higher harmonics, and give the estimations of the effect. We shortly discuss the parametric excitation.
In Section \ref{sec:Shapiro-res} we study the mechanical response at the Shapiro steps  in the regime where the a.c. driving frequency matches the resonant frequency and present the mechanisms and peculiarities of this response. In Section  \ref{sec:Shapiro-non-res} we consider the non-resonant driving that appears to efficiently excite the mechanical motion in the regime of Shapiro steps.
 Our concluding remarks are presented in Section \ref{concl}.

\section{The setup}
\label{prop}
\begin{figure}
		\includegraphics[width=0.9\columnwidth]{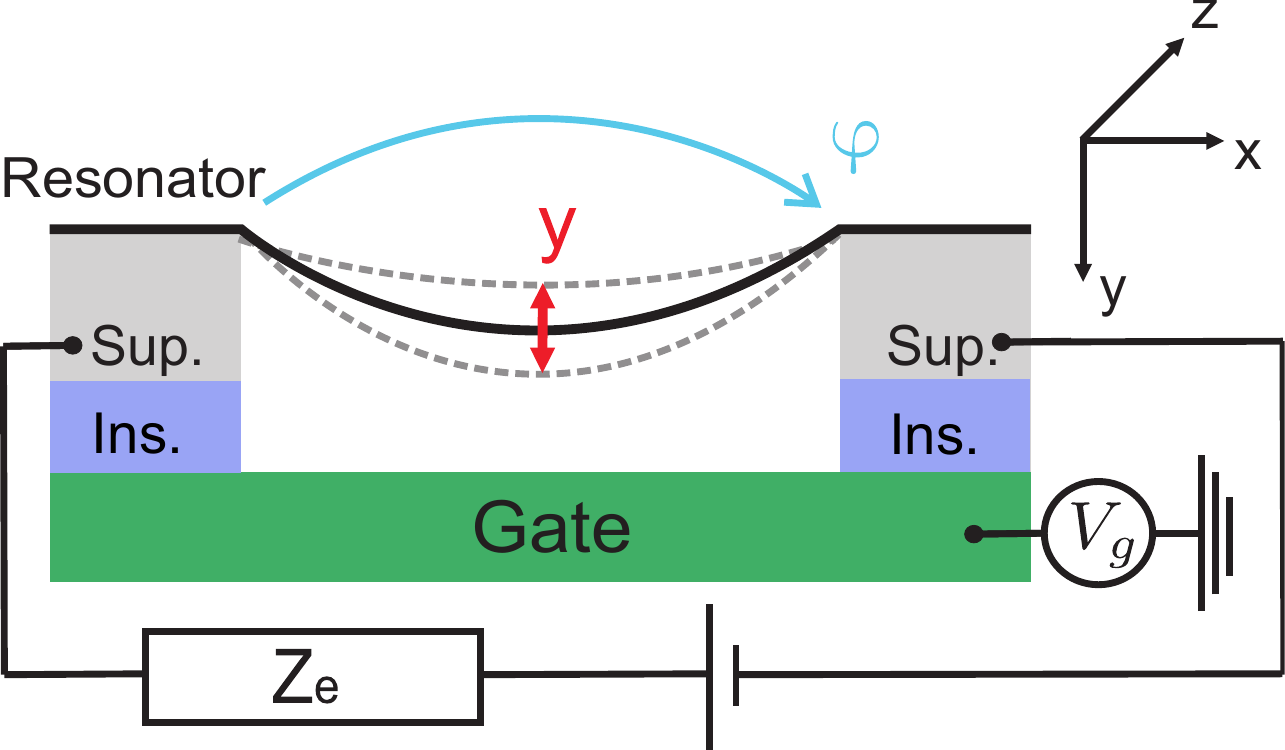}
	\caption{(Color online.) The setup. The sketch presents the mechanical resonator realized as a CNT suspended over two superconducting leads isolated from the back gate electrode. The CNT center is displaced in $y$-direction by an electrostatic force produced by the gate voltage. The superconducting leads are parts of an electrical circuit characterized by an impedance $Z_e$. The setup can be biased by either voltage or current source. }
	\label{CNT}
\end{figure}

The setup under consideration is sketched in Figure \ref{CNT} where we concentrate on a case where both Josephson junction and mechanical resonator are realized with the same single CNT. Even in this case, the coupling between mechanical and electrical degrees of freedom is relatively weak. This permits us to describe the electrical and mechanical aspects of the setup separately. We provide such description in this Section, while in the next Section we concentrate on the coupling.

\subsection{Electrical Setup}
\label{sec:ele-setup}
We consider a conducting link between two superconducting leads (the CNT in Fig. 1). In general, the current flowing in this  junction is a  complicated non-linear and time-delayed function of superconducting phase difference between the leads. However we assume that in the relevant frequency range the current response of the junction is superconducting and instant.

The junction is included into an external electric circuit and the voltage drop at the junction is related to the time derivative of Josephson phase, $\dot{\varphi}=2eV/\hbar$. In general, a circuit that connects the leads can be described by a complex frequency-dependent impedance $Z_e(\omega)$ in series with a voltage source $V_b$. We typically assume that  $Z_e$ by far exceeds the typical junction impedance at low frequencies while
at frequency scale of Josephson generation frequency $\omega \simeq eV/\hbar$,  $Z_e(\omega)$ is negligible in comparison with the junction response. In this case, the junction is current-biased at low frequencies with $I_b = V_b/Z_e(0)$ and voltage-biased at Josephson frequencies.
While this scheme looks different from the traditional RSJ model where the external impedance is connected in parallel and the junction is current-biased, it is equivalent to a generalized RSJ upon transforming the impedance and the voltage source. For instance, the linear part of possible quasiparticle response of the junction can be incorporated into $Z_e(\omega)$. 

In addition, the junction is affected by the gate electrode biased by voltage source $V_g$. The bias and gate circuits are disconnected at zero frequency. At finite frequency, there is  a cross-talk between the circuits which is difficult to eliminate or even characterize in realistic experimental circumstances. We account for that by correlating a.c. parts of the voltage sources $V_{b,g}$.
For instance, if the gate voltage consists of a d.c. part and a harmonic signal at frequency $\Omega$,
$V_g(t)=V_{g0}+ \tilde{V}_g\cos(\Omega t + \chi)$, the bias voltage source should also oscillate at the same frequency, $V_b(t)=V_{b0}+ \tilde{V}_{b}\cos(\Omega t)$. The ratio of two a.c. amplitudes and their mutual phase shift $\chi$ is determined by details of the crosstalk. We will show below that the interference of these two a.c. signals may strongly affect the d.c. currents in the junction.

The superconducting current is determined by the instant phase difference, $I(t) = I(\varphi(t))$. In this case, it is related to the Josephson energy $E_j$ of the junction,
\begin{align}
I_s=&\ (2e/\hbar) \partial E_j(\varphi)/\partial \varphi\quad.\label{IsEj} 
\end{align}
It is essential for us that the Josephson energy is not only a function of phase difference
but also depends on the gate voltage through the charge $q = C_g V_g$ induced in the resonator, $E_j = E_j(q,\varphi)$.
 
For a nanotube device, the origin of this charge sensitivity is (weak) Coulomb blockade of electrons in the middle of the nanotube. The nanotube can be viewed as  two junctions in series, those being formed at contact with metallic leads. If the conductance of the junctions is smaller or comparable with the conductance quantum $G_Q \equiv e^2/(\pi\hbar)\approx 7.75\times 10^{-5}\ \Omega^{-1}$ Coulomb interactions become important and set a quasiperiodic dependence of Josephson energy on $q$ with a period $2e$. This corresponds to charge quantization in the middle of the nanotube. We routinely observe the quasi-periodic modulation of superconducting currents in fabricated nanotube devices. The modulation can be tuned by changing the gate voltage at scale of $q \simeq 10-100 e$ from values of the order of $1$ to several per cent. Big modulation and well-developed Coulomb blockade require big junction resistances, this strongly suppresses the superconducting current. It is therefore advantageous to have intermediate resistances $R \simeq G^{-1}_Q$. At $R = 5\ k\Omega$ we typically observe 30\% modulation.

The superconducting current is a periodic function of the phase $I_s(\varphi)=I_s(\varphi+2\pi)$ and therefore can be expanded in harmonics as \cite{Likharev}: 
\begin{align}
I_s(q,\varphi) = &\ I_1(q)\sin(\varphi) + \displaystyle\sum^{\infty}_{n=2}I_{n}(q)\sin(n\varphi)\quad,\label{genIs}
\end{align}
If one neglects all harmonics except the first one, $I_1$ gives the critical current of the Josephson junction. We will typically assume this, and will mention the effect of higher harmonics only if it is crucial.


\subsection{Mechanical Setup}
\label{mechres}
Mechanical resonators can be realized in a variety of ways \cite{ways}. In many cases the adequate description of the resonator can be achieved with a minimum model that accounts for excitations of a single resonator mode, neglecting coupling to any other modes. 
The minimum model  is given by the following equation of motion for a displacement variable $y$: 
\begin{align}
\ddot{y}+\Gamma\dot{y}+\omega_0^2y -\alpha y^2 -\beta y^3=&\  F(t)/M \quad. \label{yt}
\end{align}
Here  $F(t)$ is the time-dependent driving force, $M$ is the effective mass corresponding to the mode, $\omega_0$ stands for the resonant frequency, $\Gamma \ll \omega_0$ is the damping rate, and  $\beta$ is the parameter describing  the leading cubic non-linearity \cite{Holmes}. The cubic non-linearity provides the important restriction on the magnitude of the displacement at resonant frequency as a reaction on resonant force. We also keep the second-order non-linearity $\alpha$. Although it is not important in analysis of the reaction at resonant force, it describes the shift of the resonant frequency due to constant force.

Our preferable realization of mechanical resonator is a suspended  ultra-clean CNT \cite{McEuwen, Herre, Gary} that demonstrates best quality factors observed so far($Q\equiv \omega_0/\Gamma \simeq 10^5$). In this Subsection we review the parameters of the minimal model for this realization. 
In the setup shown in Figure \ref{CNT}, the nanotube displacement from equilibrium position  and  the driving force are in the $y$-direction towards the gate, that is, perpendicular to the nanotube axis.
The mechanical variable $y(t)$ is the displacement of the mid-point of the nanotube. 

In the case of a CNT, the adequate model of mechanical properties involves a suspended thin cylindrical rod clamped at both ends where the nanotube touches the metal leads. The parameters are the rod length $L$, the cylinder radius $r$, and the tube cross section area $S$. 
In our experiments, $L\simeq 0.3..0.5\ \mu m$, $r \simeq 1\ nm$, and $S = 2\pi r a \simeq 2.1\ nm^2$ for a single-wall nanotube, $a \simeq 0.34\ nm$ being the layer spacing in graphite.   
The relevant elastomechanical material constants, carbon Young's modulus $E$ and graphite density $\rho$ are estimated as $E \simeq 10^{12}\ J/m^3$ \cite{Young} and $\rho \simeq 2.2\ g/cm^3$. 
The bending modes of the rod and their complete dynamics are described by the Euler-Bernoulli equation of motion \cite{Landau, Yaroslav}. 
 
We concentrate on the lowest frequency bending mode that has no nodes in the rod and therefore is easy to excite.
  The resonant frequency can be tuned by "tightening" the tube, that is, changing the elastic tension. This is achieved by applying a sufficiently big d.c. gate voltage $V_{g0}$. The resulting electrostatic force strives to elongate the nanotube, thus producing the tension. In such a way, the resonant frequency can be increased by a factor of three in comparison with that of the "loose" nanotube. For estimations, we concentrate on the case of untightened rod. In this case, the resonance frequency corresponding to the lowest CNT bending mode can be estimated in terms of the bending spring constant and the carbon mass density $\omega_0 \simeq 22.4 \sqrt{EI/\rho S}L^{-2}$ \cite{Landau}, where $I=Sr^2$ is the principal moment of inertia. In our devices of length $L= 0.3..0.5$ $\mu m$, the frequency is $\omega_0/2\pi \simeq 0.30..0.84$ GHz, similar to frequencies reported in \cite{Gary}. 
The effective force is evaluated using the eigenfunction of the mode $\xi(x) \equiv y(x,t)/y(t)$, $F=\int_0^L dx f(x)\xi(x)$, $f(x)$ being the force per unit length. For electrostatic forces, an ad-hoc assumption is that the force distribution is uniform, so the total force is $F_{t} =f L$. In this case, $F \simeq 0.53 F_{t}$.
The effective mass is given by 
$M = \rho S\int_0^L dx \xi^2(x)$, $M \simeq 0.41 \rho SL\simeq 4.1..6.8\times 10^{-22}$ kg.   
The cubic non-linearity originates from the tension produced by the nanotube displacement, the corresponding parameter can be estimated as $\beta\simeq 40\ ES/ML^3\simeq\omega^2_0/r^2\simeq 1.8..5.5\ \text{GHz}^2nm^{-2}$, assuming uniform distribution of force along the length of the rod. The second-order non-linearity $\alpha$ vanishes for untightened straight rod for symmetry reasons. It however appears if the rod is tightened such that its frequency 
change with respect to the untightened value $\omega_0$ is of the order of $\omega_0$. In this case, the non-linearity is obtained as $\alpha = 3 \beta y_0$, $y_0$ being the equilibrium displacement induced by the tightening. 
 
If $F(t)$ oscillates at frequency $\omega$ close to the resonant frequency,  Eq. (\ref{yt}) can be solved in resonant approximation for the complex amplitude  $\tilde{y}$:
\begin{align}
\tilde{y} =&\ \frac{\tilde{F}}{2M\omega_0}\frac{-1}{\nu+i\Gamma/2+(\beta\rq{}/2\omega_0)|y_{\omega}|^2} \quad,\label{y}
\end{align}
with $\tilde{F}$ being the complex force amplitude. 
Here we introduce the detuning $\nu\equiv \omega-\omega_0$ implying that $|\nu|\lesssim \omega_0$. We also introduce the Duffing parameter $\beta\rq{}=\beta+\alpha^2/\omega_0^2\simeq \omega_0^2/r^2$ corresponding to the amplitude-dependent frequency shift. In our experiments, we estimate $\beta\rq{}\simeq 3.6..11\ \text{GHz}^2nm^{-2}$.

\begin{figure}
		\includegraphics[width=0.9\columnwidth]{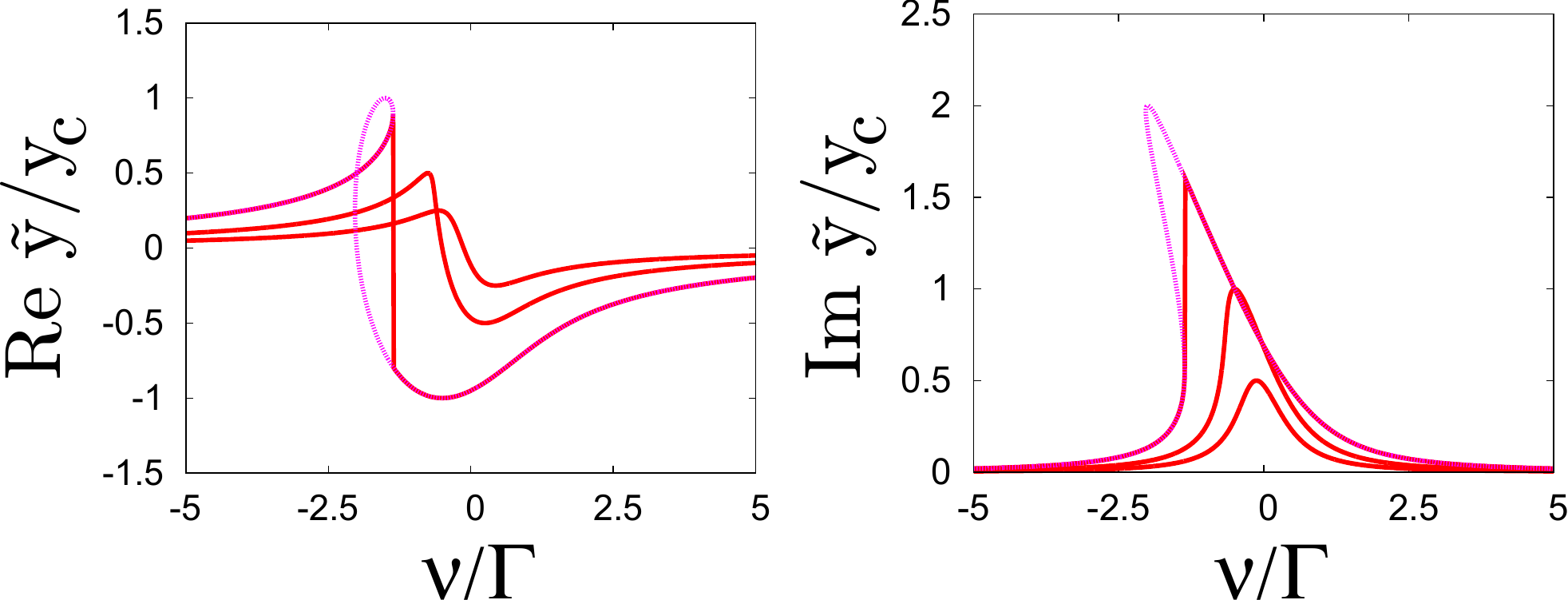}
	\caption{(Color online.) Left panel: Real part of the complex displacement amplitude  ${\rm Re}\;\tilde{y}$ versus detuning. Right panel: Imaginary part of the complex displacement amplitude  ${\rm Im}\;\tilde{y}$ versus detuning. The three curves in each panel correspond to $\tilde{F}/F_c=0.5,1,2$.}
	\label{fig:real-imaginary}
\end{figure}

We will re-write Eq. \eqref{y} in dimensionless form introducing a critical amplitude $y_c$,  $y_c=\sqrt{\omega_0\Gamma/\beta\rq{}}$. At this amplitude scale, the response of the resonator becomes a two-valued function of detuning (see Fig. \ref{fig:real-imaginary}) 
For a CNT, it can be estimated as $y^2_c\simeq r^2/Q$ which corresponds for our experiments to $y_c\simeq 3.2\ pm$. The driving force corresponding to $y_c$ is $F_c = M\beta\rq{} y^3_c= M\omega_0^2 y_c/Q$. We estimate it for a CNT $F_c\simeq 10^2\ ES (r/L)^3Q^{-3/2}\simeq 1.2\times 10^{-18}$ N.

The dimensionless form of Eq. (\ref{y}) is:
\begin{align}
\frac{\tilde{y}}{y_c} =\frac{\tilde{F}}{F_c} R\left(\frac{\nu}{\Gamma},\frac{\tilde{F}}{F_c}\right);\;
R(a,b) = &\ \frac{-1}{2a+|b|^2|R(a,b)|^2+i} 
\label{yc}
\end{align} 
Here we have introduced a dimensionless complex response function $R(\nu,\tilde{F})$. In the linear regime $|\tilde{F}|\ll F_c$ its dependence on $\tilde{F}$ can be neglected: $R(a,b)=(2a+i)^{-1}$.

Figure \ref{fig:real-imaginary} shows the real and imaginary parts of $\tilde{y}$ as a function of detuning for three values of the driving force amplitude that correspond to quasi-linear, critical and bistable regimes.

\section{Coupling and non-linearities}
\label{Fe}

In this Section, we analyze the coupling between mechanical and electrical degrees of freedom, with emphasis on phase-dependent mechanical force.
Since we prove that this force emerges from charging effect, we will start with a detailed discussion of electrostatic energy in the setup, and express  the forces and superconducting current  in terms of  the induced charge. By doing so, we assume that the typical time of charge equilibration is much shorter than the typical time scale $\omega_0^{-1}$ of the mechanical motion.  We compare the electrostatic phase-dependent force with Lorentz force proposed in \cite{Sonne}. We derive the coupled equations of motion governing the Josephson and mechanical dynamics and identify the dominant source of non-linear behavior. 

\subsection{Electrostatic energy}
The junctions connecting the middle of the nanotube to the leads have intermediate resistance so that the middle of the nanotube forms a Coulomb island that is neither isolated from nor ideally connected to the leads. While this situation is difficult to quantify from a microscopic calculation, it can be completely analyzed at phenomenological level. The consideration below is just a case of elementary non-linear electrostatics and is similar to the discussion in \cite{Yaroslav}. However, it contains some important and less obvious details so we choose to present it at a comprehensive level.

To start with, let us assume that the capacitance to the gate is vanishingly small while $V_g$ is diverging such that the charge induced to the Coulomb island by the gate 
$q= C_g V_g$ tends to a constant limit. A part of the ground-state energy of the setup, $E_c(q)$ does depend on $q$. General Coulomb-blockade considerations \cite{YuliBook} imply that this part is a (quasi) periodic function of $q$ with a period of $2e$. In the limit of full isolation, for instance, this energy is piecewise parabolic,
$E_c(q) = E_C {\rm min}_N\left(N-q/2e\right)^2$, $N$ being an integer number of extra Cooper pairs stored in the island. In general, it is a smooth function of $q$ and may depend on the superconducting phase difference $\varphi$ and, in principle, on mechanical displacement $y$.
This energy results in a non-zero electrostatic potential difference between the island and the leads, $V(q) = - \partial E_c(q)/\partial q$. 

Let us now turn to finite $C_g$ and therefore finite $V_g$ that is the potential difference between the leads and the gate electrode. Since this is not the potential difference between the island and the gate anymore, the induced charge $q$ is not equal to $C_gV_g$. Rather, it is determined from the voltage division in a nonlinear capacitance network, or, equivalently, from the minimization of the total electrostatic energy with respect to $q$,
\begin{align}
E ={\rm min}_q \left( E_c(q)+\frac{q^2}{2C_g} - qV_g \right)\quad . \label{Jq}
\end{align}
The charge is then found from the condition of the minimum,
\begin{align}
\label{eq:for_charge}
q= C_g V_g -  C_g\frac{\partial E_c(q)}{\partial q}\quad. 
\end{align}
There are two implicit dependences in this equation that distinguish it from pure electrostatics, and that we make explicit now. First of all, the electrostatic energy depends on the mechanical displacement of the nanotube. Geometric considerations suggest that this dependence can be ascribed to $C_g$: Indeed, the modification of capacitance to the gate is linear in $y$, $C_g \to  C_g  + \frac{d C_g}{d y} y$ while the modification of $E_c$ is expected to be $\propto y^2$. Second, the electrostatic energy depends on the superconducting phase difference: indeed, the Josephson energy is just the phase-dependent part of $E_c$, $E_c(q,\varphi) = \bar{E}_c(q)+E_j(q,\varphi)$. The electrostatic charge $q$ depends both on displacement $y$ and on superconducting phase $\varphi$.

To single out these contributions, we assume that i. the voltage between the middle of the nanotube is smaller than the gate voltage, $\partial E_c/\partial q\ll V_g$, this is fulfilled if the induced charge $q\gg e$, i.e., in any practical setup; ii. the mechanical displacement is small in comparison with the distance to the gate, $y\ll C_g\left(\frac{dC_g}{dy}\right)^{-1}\simeq L_g$ (in our experiments $L_g\simeq L\gg y$). With this, we linearize Eq. (\ref{eq:for_charge}) with respect to the Josephson energy   and mechanical displacement to arrive at ($q_0 \equiv C_g V_g$)
\begin{align}
q = q_0 + V_g\frac{dC_g}{dy}y - C_g\frac{\partial E_j}{\partial q}\left(q,\varphi\right)
\label{linset1}
\end{align}
The first term is the common expression for the gate-induced charge while the second and the third are the corrections of interest. At the moment , we keep $q$ in the argument of
$E_j$, although $q \approx q_0$. The point is that the Josephson energy is sensitive
to variations of $q$ of the order of $e$, and ($q-q_0$) can in principle be of this order. 
Since $y \ll L_g$, we may disregard the possible $y$ dependence of $dC_g/dy$.

\subsection{Forces}
\label{Jforces}

The mechanical resonator is affected by the electrostatic force $F=-\partial E/\partial y$:
\begin{align}
\label{eq:for_force}
F = \frac{dC_g}{dy} \frac{q^2}{2C_g^2}\quad. 
\end{align}
It is convenient to distinguish three separate contributions to the total force: the static force, the gate driving force and the phase-dependent Josephson force. 

The static force is produced by the d.c. gate voltage. Its magnitude is given by $F_{\rm static}=(dC_g/dy)V_{g0}^2/2$, corresponding to the first dominating term in \eqref{linset1}.The effect of the static force is to pull on the resonator, thereby tuning its resonance frequency\cite{Gary}.
Since it is stationary, it does not excite the oscillations.

The a.c. gate driving force arises due to the a.c. modulation of the gate voltage and is given by $F_g = (d C_g/dy)V_{g0} \tilde{V}_{g}\simeq (q_0/e)(e\tilde{V}_g/L_g)$.

The phase-dependent Josephson force, not discussed in previous literature, comes about the product of the first and third term in 
(\ref{linset1}),
\begin{align}
F_j = &\ -\frac{dC_g}{dy}V_{g0}\frac{\partial E_j(q, \varphi)}{\partial q}\quad , \label{Fj1}
\end{align}
In fact it is similar to the gate driving force, with $\tilde{V}_g$ replaced by the phase-dependent voltage arising in the capacitive network, $\partial E_j(q,\varphi)/\partial q$. In contrast to the gate driving force, the time dependence of the Josephson force is determined by the phase dynamics  rather than the external modulation of the gate voltage.

The scale of the Josephson force is $\bar{F}_j\simeq (q/e)(E_j/L_g)$, where we assume $\partial E_j/\partial q \simeq E_j/e$, which is true for intermediate contact conductances $\simeq G_Q$. It can be compared to the scale of a.c. gate driving force, $F_j/F_g\simeq E_j/e\tilde{V}_g$. For sufficiently low a.c. driving amplitude $e\tilde{V}_g\ll E_j$ the Josephson force dominates $F_g\ll F_j$.

\subsection{Lorentz force}
The Josephson force explained above arises from  the combined effect of charge sensitivity of the Josephson coupling and the capacitive coupling to the gate electrode. 
The alternative mechanism of generating a $\varphi$-dependent force was recently proposed by G. Sonne {\textit et al} \cite{Sonne}.
This force is of Lorentz type arising from the interaction of the $\varphi$-dependent superconducting current with an external magnetic field $\vec{B}$ applied in the perpendicular direction (for the setup of Figure \ref{CNT}, along the $z$-axis.)
This mechanism does not require the presence of a gate.
 
Let us compare the Lorentz force and the electrostatic Josephson force. The Lorentz force is in $y$-direction, that is, perpendicular to both $\vec{B}$ and the supercurrent, $F_{\rm B}= L |\vec{B}| I_s$. It is natural to express $F_j$ in terms of the electric field $|\vec{E}|=(dC_g/dy) V_g/C_g$ produced by the gate electrode. For estimates, we assume $L \simeq L_g$ and $(dC_g/dy)L/C_g\simeq L/L_g\simeq 1$. This yields
\begin{align}
\frac{F_{\rm B}}{F_j} \simeq &\ \frac{c |\vec{B}|}{|\vec{E}|} \alpha\quad,
\end{align}
where $c \simeq 3\times10^8$ $m/s$ is the speed of light  and $\alpha = e^2/4\pi\epsilon_0\hbar c \simeq 1/137$ is the fine structure constant. 

Typical magnetic fields used in experiments are $|\vec{B}|\simeq 1$ $T$. They are limited from above by the critical fields of the superconducting leads. The typical electric fields are  $|\vec{E}|\simeq 10^7$ $V/m$. This corresponds to a potential drop of $V_g\simeq 10$ $V$ over a distance of $L_g\simeq 0.5$ $\mu m$. For these  values $F_{\rm B}/F_j\simeq 10\alpha \ll 1$ suggesting that the Josephson force dominates. Therefore in  the rest of the paper we will disregard the Lorentz force.

If one imagines a ballistic nanotube, the Josephson coupling is not affected by the induced charge. In this case the Lorentz force would be the only superconducting phase-dependent driving mechanism. However, the ideally ballistic nanotubes have not been realized experimentally.

\subsection{Analysis of non-linearities}

Let us bring together three coupled equations
governing the dynamics of the setup
 \begin{align}
\frac{V_b(\omega)}{Z_e(\omega)}+i\frac{\hbar\omega}{2e}\frac{\varphi(\omega)}{Z_e(\omega)}= &\ \left(\tilde{I}_s(q,\varphi)\right)_\omega \ ,\label{set2}\\
\ddot{y}+\Gamma\dot{y}+\omega_0^2y +\alpha y^2 -\beta y^3=&\ M^{-1}\frac{dC_g}{dy} \frac{q^2}{2C_g^2}\:.\label{set3}\\
 q_0 + \frac{q_0}{C_g}\frac{dC_g}{dy}y - C_g\frac{\partial E_j}{\partial q}\left(q,\varphi\right)=&q \label{set1}
\end{align}
The first equation describes the dynamics of superconducting phase difference $\varphi(t)$ and is obtained by applying Kirchhoff's laws to the circuit. The second equation is for the mechanical displacement $y(t)$ where we substitute the electrostatic force (\ref{eq:for_force}). The induced charge $q$ enters both equations, and at the same time is defined by the third equation, that is, its value depends both on $\varphi$ and $y$. Therefore the equations are coupled.  

We wish to simplify these equations under experimentally relevant assumptions. For this,
we shall analyze the relative importance of different non-linearities in the coupling.
There are natural non-linearity scales for all three variables, $\varphi \simeq 2\pi$, $q \simeq e$, $y \simeq y_c \simeq \sqrt{\omega_0 \Gamma/\beta'}$. This could change if the coupling is sufficiently strong. For instance, the displacement may cause the variation of phase that is subject to Josephson non-linearity. The resulting variation of phase would produce the non-linear variation of $q$, this will result in non-linear feedback on $y$. This could in principle cause a non-linear scale of $y$ to be smaller than $y_c$.
So first of all we shall quantify the coupling between electrical and mechanical variables by comparing the non-linear terms in the mechanical force resulting from the coupling with those coming from the intrinsic  non-linearities characterized by $\alpha$ and $\beta$.

The conclusion of this Subsection is that the mechanical non-linearity is the dominant one.
We prove this with a rather involved reasoning given below.

For the estimations, it is convenient to introduce the following dimensionless parameters:
\begin{eqnarray*}
A_j =&\ C_g\frac{\partial^2E_j}{\partial q^2}\simeq \frac{E_j}{E_C} ,\  E_C \equiv \frac{e^2}{C_g}\\
B_j =&\ \frac{2e^2Z_e}{\hbar\omega_0} \frac{\partial\tilde{I}_s}{\partial q}\simeq G_QZ_e\frac{E_j}{\hbar\omega_0}
\end{eqnarray*}
For estimations, we assume that $A_j,B_j$ are either small or of the order of $1$.
This assumption is valid for $A_j$; it compares the Josephson energy to the charging energy under conditions of well-developed Coulomb blockade. The parameter $B_j$ is a coefficient of Josephson feedback at high frequency and depends on the details of the external circuit via the impedance $Z_e$. Unless a special effort is made to increase the circuit impedance at high frequency, $B_j$ will not be big. 
 
Given a  variation of displacement $\delta y$ we estimate  the linear responses of the charge $\delta q$ and the superconducting phase $\delta \varphi$ on $\delta y$ using Eqs. (\ref{set1}) and (\ref{set2}) as 
\begin{align}
\delta q \left(1+A_j+A_jB_j\right) = &\ q_0\frac{1}{C_g}\frac{dC_g}{dy}\delta y \simeq q_0 \frac{\delta y}{L_g}\label{dq}\\ 
\delta\varphi\left(1+A_j+A_jB_j\right)= &\ B_j \frac{q_0}{e}\frac{1}{C_g}\frac{dC_g}{dy}\delta y \simeq B_j \frac{q_0}{e} \frac{\delta y}{L_g} \label{dphi}
\end{align}
Assuming $A_j,B_j\lesssim 1$ the linear responses can be estimated as
$$
\delta q \simeq q_0 \frac{\delta y}{L_g},\quad \delta\varphi\simeq B_j \frac{q_0}{e} \frac{\delta y}{L_g}\quad.
$$
We use this to find a scale of $\delta y_c$ for which the responses of charge and superconducting phase may become comparable with the scales of their intrinsic non-linearities $\delta q \simeq e$ and $\delta \varphi\simeq 2\pi$. 
We conclude that $\delta y_c \simeq L_g (e/q_0)$. 
Comparison with the scale $y_c \simeq r/\sqrt{Q}$ of the mechanical  non-linearity yields
\begin{equation}
\frac{\delta y_c}{y_c} \simeq \frac{e}{q_0}\frac{L_g}{r}\sqrt{Q}
\end{equation}
Two last factors in this expression are big, while the first one can be small.
We estimate the biggest $q_0$ from the condition that $\omega_0$ is changed significantly by applying the gate voltage, that is, the stationary displacement $y_0 \simeq r$. This yields 
\begin{equation}
\frac{q_0}{e} \simeq r^2 L^{1/2} a^{3/2},
\label{eq:q0-estimation}
\end{equation}
$a$ being atomic scale, $q_0 \simeq 10^2 e$ for our devices.
The fact that the intrinsic scale of the non-linearity is smaller signals that the intrinsic non-linearity of the resonator dominates the non-linearity arising due to coupling. With this, we estimate the first two factors as $(e/q_0)(L/r) \simeq (a L/r^2)^2$. This is $\simeq 10$ for our geometries and we conclude that $\delta y/y_c \gg 1$ for any $Q >1$. 

This proves that the dynamics of charge and phase is linear in $y$ provided our estimations of mechanical non-linearities $\alpha,\beta$ hold. 
We still need to show that the coupling to Josephson junction does not change these nonlinearities significantly.

So we estimate  the quadratic and cubic non-linearities of the mechanical force due to coupling. First we find the quadratic and cubic variations of charge with respect to displacement using Eq. (\ref{set1}).
\begin{align}
\delta q^{(2)} = &\ C_g\frac{\partial^3E_j}{\partial q^3}(\delta q)^2 \simeq e A_j \left(\frac{q_0}{e} \frac{\delta y}{L_g}\right)^2 \quad, \label{dq2}\\ 
\delta q^{(3)} = &\ C_g\frac{\partial^4E_j}{\partial q^4}(\delta q)^3 \simeq e A_j \left(\frac{q_0}{e} \frac{\delta y}{L_g}\right)^3 \quad. \label{dq3}
\end{align}
We can now estimate the terms in the  mechanical force that are 
quadratic and cubic in $\delta y$.
\begin{align}
\delta F^{(2)} = &\ \frac{dC_g}{dy} \frac{q_0^2}{2C_g^2}\left(2\left(\frac{\delta q}{q_0}\right)^2+\frac{\delta q^{(2)}}{q_0}\right)\notag\\
\simeq &\ F_{\rm stationary} \left(1+A_j\frac{q_0}{e}\right)\left(\frac{\delta y}{L_g}\right)^2\quad, \label{dF2}\\
\delta F^{(3)} = & \ \frac{dC_g}{dy} \frac{q_0^2}{2C_g^2}\left(3\frac{\delta q}{q_0}\frac{\delta q^{(2)}}{q_0}+\frac{\delta q^{(3)}}{q_0}\right)\notag\\
\simeq &\ F_{\rm stationary} A_j\left(\frac{q_0}{e}\right)^2\left(\frac{\delta y}{L_g}\right)^3 \quad.\label{dF3}
\end{align}

 We compare these terms with the intrinsic non-linearities. The second order term $\delta F^{(2)}$ needs to be compared with the mechanical quadratic non-linearity $M \alpha \delta y^2$. Assuming the static displacement of the order of CNT radius, $y_0\simeq r$, we estimate $M\alpha \simeq M\beta/r \simeq M\omega_0^2/r$.  
\begin{equation}
\frac{\delta F^{(2)}}{M \alpha \delta y^2}\simeq \left(1+A_j\frac{q_0}{e}\right)\frac{r^2}{L^2_g}\ll 1\quad . 
\end{equation}
 Here we use the estimation $(q_0/e)(r/L_g) \ll 1$,  $(q_0/e)(r/L_g) \simeq 0.1$ for typical CNT geometries. (see Eq. \eqref{eq:q0-estimation}).
 The third order term $\delta F^{(3)}$ needs be compared with the third-order non-linearity $M\beta\delta y^3$. This yields 
\begin{equation}
\frac{\delta F^{(3)}}{M\beta\delta y^3}\simeq A_j\frac{r^3}{L^3_g}\ll 1\quad . 
\end{equation}

To summarize, we proved that the  non-linear scales correspond to $\varphi \simeq 2\pi$, $q \simeq e$, $y \simeq y_c \simeq \sqrt{\omega_0 \Gamma/\beta'}$ and that for a CNT resonator the intrinsic mechanical non-linearities dominate the non-linearities arising from coupling. This permits a simplification of the dynamical equations. We may linearize the terms describing the coupling of mechanical displacement and electricity, thus separating Josephson and mechanical non-linearities. 

\subsection{Workflow}

This sets the following workflow.
\begin{itemize}
 \item {\ }At given a.c. and d.c. bias and gate voltages we solve for Josephson dynamics neglecting the mechanical coupling and setting $q =q_0(t)$. We find $I(t)$ and $\varphi(t)$.  
 Using these, we compute the Josephson force $F_j$ given by Eq. \eqref{Fj1}. 
\item{\ } We solve the non-linear mechanical equation 
\begin{equation}
M(\ddot{y}+\Gamma\dot{y}+\omega_0^2y +\alpha y^2 -\beta y^3)=F_{\rm st} +F_g+F_j\:.\label{simpleset3}
\end{equation}
to find $y(t)$. We are mostly interested in a part that oscillates with frequency $\simeq\omega_0$.
This may be excited by both $F_j$ and $F_g$. 
\item{ \ }In most cases, we are interested in a (d.c.) current response on the mechanical motion, the mechanical response.
It arises due to direct modulation of charge by the mechanical displacement,
\begin{equation}
\tilde{I}_\text{mh}(t) =  \frac{2e}{\hbar} \frac{\partial^2 E_j}{\partial \varphi \partial q} \frac{d C_g}{dy} V_{g0} y(t)
\label{eq:workflow-tilde-Im}
\end{equation}
in the first order in $y(t)$. 
We will mostly concentrate on the situation when the displacement oscillates at 
the resonant frequency while a d.c. component of $\tilde{I}_\text{mh}$ is of interest.
The d.c. signal them arises from the rectification of $y(t)$ by an oscillating part of
$\partial I(\phi)/\partial(q/e)$, that we call the detecting current.
    
\item{\ } If we can neglect the feedback in Josephson dynamics, we are done, since
the response is given directly by $\tilde{I}_\text{mh}$. Otherwise, we linearize the Josephson dynamics to determine the response of the superconducting phase on the mechanical displacement found, $\varphi_\text{mh}(t)$
$$
\varphi_\text{mh}(t) = \frac{\hbar}{2e} \int^t dt' dt'' Z(t',t'')  \tilde{I}_\text{mh}(t'')
$$
Here the kernel $Z(t,t')$ represents the combined linear impedance of the external circuit and the junction.

\item{\ } Taking this into account, we obtain the current response sought:
\begin{equation}
I_\text{mh}= \tilde{I}_\text{mh}+ \frac{2e}{\hbar} \frac{\partial^2 E_j}{\partial \varphi^2} \varphi_\text{mh}
\label{eq:workflow-Im}
\end{equation}

The first term is the direct modulation of the current by the charge induced by the mechanical displacement while the second one is a feedback of Josephson junction by means of $\varphi_\text{mh}$.

\end{itemize}

The mechanical response is thus typically a small correction to the maximum superconducting current.
We can estimate it at maximum taking the 
$$
\frac{I_m}{I_c} \simeq \frac{q_0}{e} \frac{y_c}{L} \simeq \frac{q_0 r }{\sqrt{Q} e L} 
\simeq 10^{-3}
$$
(We remind that $(q_0/e)(r/L) \simeq 0.1$, $Q \simeq 10^{5}$ for our devices).
Perhaps unexpectedly, the typical response becomes smaller upon increasing the quality factor. The reason  for this
is clear: the maximum displacement becomes smaller. However, the large $Q$ results in sharp frequency dependence of the response making it easier to identify.
We thus concentrate on this dependence.

\subsection{Parameters}

Let us specify the values of parameters employed. We choose these values such that they closely match those of a typical experimental realization of our setup.
\begin{align}
\label{eq:parameter-set}
\ & \text{Junction critical current, } I_c:\ 1.0\times 10^{-8}\ \text{A}; \\
\ & \text{Josephson energy, } E_j:\ 2.1\times 10^{-5}\ \text{eV}; \nonumber\\
\ & \text{Static gate voltage, } {V}_{g0}:\ 1.0\ \text{V}; \nonumber\\
\ & \text{Static charge on the resonator , } q_0/e: \  100; \nonumber\\
\ & \text{Josephson force, } F_j: \  1.1\times 10^{-15}\ \text{N}; \nonumber\\
\ & \text{Resonator length and distance to gate, } L=L_g: \  0.3\ \mu\text{m}; \nonumber\\
\ & \text{Resonator mass, } M: \  4.1\times 10^{-22}\ \text{kg}; \nonumber\\
\ & \text{Resonance frequency, } \omega_0/2\pi: \  0.84\ \text{GHz}; \nonumber\\
\ & \text{Quality factor, } Q: \  1.4\times 10^{5}; \nonumber\\
\ & \text{Quadratic non-linearity, } \alpha: \  5.5\ \text{GHz}^{2}\text{nm}^{-1};\nonumber\\
\ & \text{Cubic non-linearity, } \beta: \  5.5\ \text{GHz}^{2}\text{nm}^{-2}; \nonumber\\
\ & \text{Scale of maximum displacement, } y_c: \  3.2\ \text{pm}; \nonumber\\
\ & \text{Mechanical force scale, } F_c: \  1.2\times 10^{-18}\ \text{N}.\nonumber
\end{align}

\section{Phase bias}
\label{sec:phasebias}
Let us start our considerations with the junction biased with a time-independent phase $\varphi$:
such bias condition can be achieved by embedding the junction into a superconducting loop. Unfortunately, our present experimental setup does not allow measurements under these bias conditions. We present the theoretical results in hope that they will be useful for future experiments.

The simplest experimental signature of Josephson force under phase bias conditions is the {\it phase-dependent shift} of the resonant frequency. The mechanism of this shift in our situation is the mechanical non-linearity: the static Josephson force tightens or looses the nanotube resulting in the frequency change. The frequency response on the static force in our model reads 
\begin{align}
\frac{d\omega_0}{dF} = &\ \frac{\alpha}{M\omega_0^3}\simeq \frac{\omega_0}{F_{\rm static}}\quad . \label{ag}
\end{align}
so the phase-dependent frequency shift reads
\begin{align}
\Delta \omega_0(\varphi) = &\ \frac{\partial\omega_0}{\partial F}F_j(\varphi) = -\frac{\alpha}{M\omega_0^3} \frac{dC_g}{dy}V_{g0} \frac{\partial E_j(q, \varphi)}{\partial q}\nonumber\\
\simeq &\ \omega_0\frac{F_j(\varphi)}{F_{\rm static}}\simeq\omega_0\frac{E_j}{eV_{g0}}\cos(\varphi)\quad . \label{shift}
\end{align}
 The shift is clearly observable provided it exceeds the broadening $\Gamma$. The estimation gives
\begin{align}
\frac{\Delta \omega_0(\varphi)}{\Gamma} \simeq Q \frac{E_j}{eV_{g0}}\cos(\varphi) . \label{compshift}
\end{align}
For the parameter set (\ref{eq:parameter-set}), maximum value of the shift 
 \begin{align}
\left(\frac{\Delta \omega_0(\varphi)}{\Gamma}\right)_{{\rm max}}
 = 2.8, \label{value-shift}
\end{align}
this is, the shift is clearly observable.

\begin{figure}
 \includegraphics[width=0.9\columnwidth]{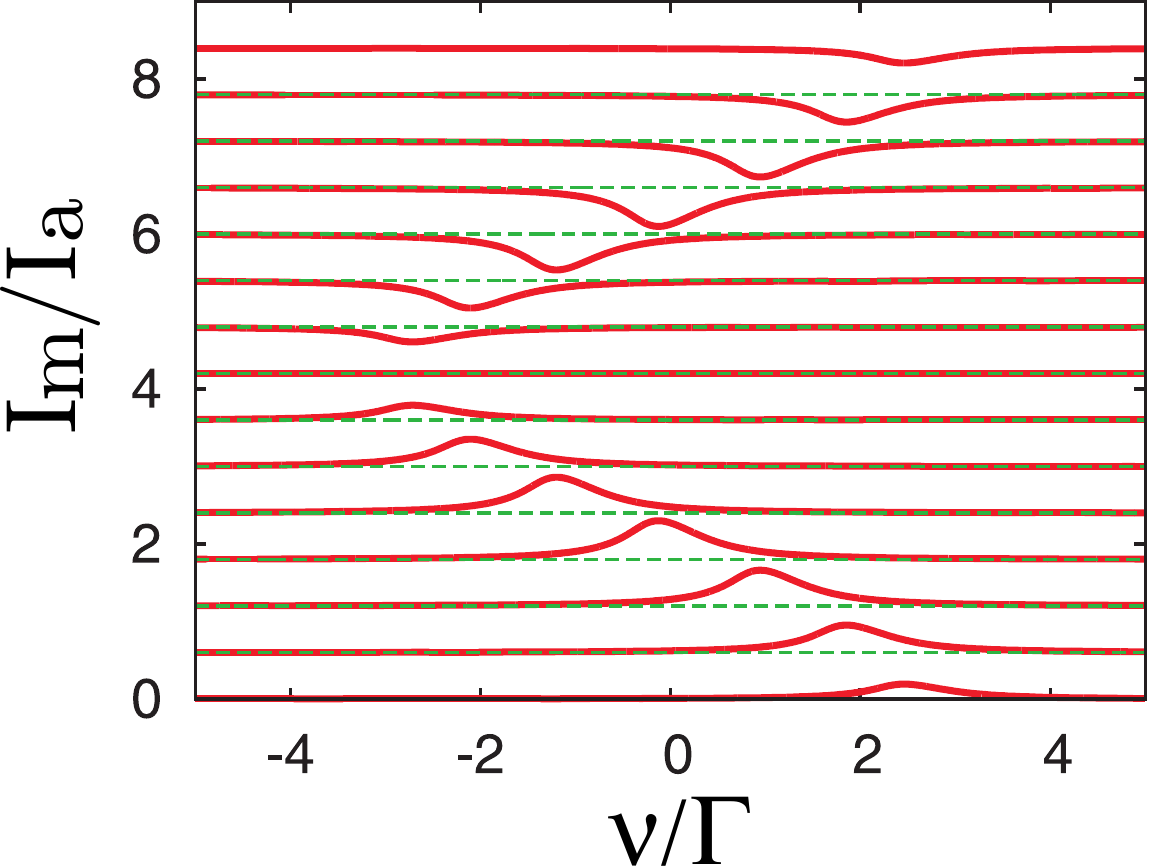}
 \caption{(Color online.) The phase-dependent frequency shift for the case of weak driving. The curves give the linear response of d.c. current in units $I_a=2e (\Delta\omega_0)_{\rm max}|\tilde{y}|^2 (M\omega_0/\hbar)$ as a function of frequency detuning for a set of phase bias values: from the lowermost to the uppermost curve the phase changes from $\varphi=\tfrac{\pi}{8}$ to $\varphi=\tfrac{15\pi}{8}$, with interval $\pi/8$. The curves are offset for clarity. Dashed lines give the positions of zero.
 \label{fig:shift}}
 \end{figure}

Let us consider an example of a mechanically-induced response under conditions of the phase bias. To excite mechanical oscillations, we apply an additional a.c. voltage to the gate that oscillates at the frequency $\Omega$ close to the resonant frequency $\omega_0$. 
\begin{align}
V_g = &\ V_{g0}+\tilde{V}_g\cos(\Omega t)\quad . \label{gatesignal}
\end{align}
Assuming $\tilde{V}_g$ to be sufficiently small to provide a linear response of the displacement, we obtain the following expression for the resonant part of the displacement:
\begin{align}
\tilde{y} =&\ \frac{\tilde{F}_g}{2M\omega_0}\frac{-1}{\nu(\varphi)+i\Gamma/2} \ ,\quad \tilde{F}_g=\frac{dC_g}{dy}V_{g0}\tilde{V}_g\ , \label{ytilde}\\
y(t) = &\ \frac{1}{2}\left(\tilde{y}e^{-i\Omega t}+\tilde{y}^*e^{i\Omega t}\right)\ ,
\end{align}
Here,  $\nu(\varphi)=\Omega-\omega_0 -\Delta\omega_0(\varphi)\ll\omega_0$ is the detuning that includes the phase dependent shift of the resonance frequency discussed above.
Owing to the mechanical non-linearity, the oscillating displacement produces a stationary displacement $y = \alpha|\tilde{y}|^2/\omega_0^2 $. This induces a stationary charge that affects the d.c. superconducting current at constant phase bias. Rather remarkably, this effect is related to the phase-dependent shift discussed above. Indeed, both are proportional to charge-dependent part of the Josephson energy and to the non-linearity coefficient $\alpha$. The resulting current response reads
\begin{align}
I_m = - 2 e \frac{\partial}{\partial \varphi} \left(\Delta \omega_0(\varphi)\right) |\tilde{y}|^2 (M\omega_0/\hbar),
\label{eq:Im-phasebias}
\end{align}
 the contribution to the current being of the order of $e \Delta \omega_0$ if we measure quantum fluctuations of displacement $\sqrt{\hbar/M\omega_0}$. The  dependence on frequency of the a.c. modulation is a Lorentzian one, as it is frequently expected
 (Fig. \ref{fig:shift}), the Lorentzian center being shifted with changing the phase.
 
 \subsection{Fano-type response}
 The above mechanism of response exploits the dominating mechanical non-linearity.
 It is proportional to $\tilde{y}^2$. Upon increase of the a.c. amplitude $\tilde{V}_g$ the
oscillating displacement saturates owing to the non-linearities. In this case, the dominating d.c. current signal can arise from the electrical non-linearity as a result of a mixing of the oscillating displacement $\tilde{y}$ and the oscillating charge $\propto \tilde{V}_g$. The resulting current is thus proportional to $\tilde{V}_g\tilde{y}$ and may exceed the contribution $\propto \tilde{y}$ provided the latter saturates.

\begin{figure}
 \includegraphics[width=0.9\columnwidth]{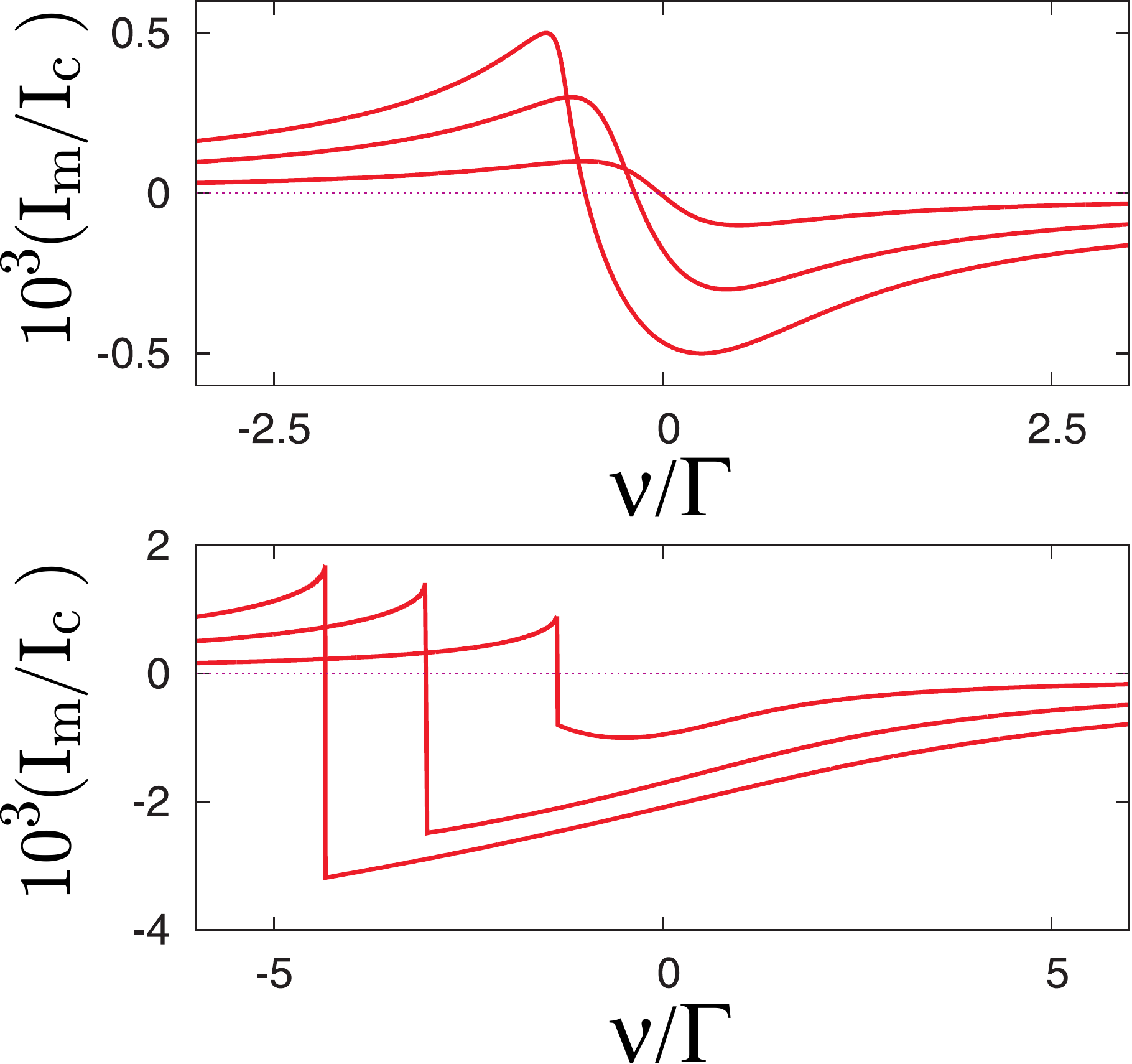}
 \caption{(Color online.) An example of Fano-type frequency-dependence of the mechanical response \eqref{eq:Fano-response}. The curves correspond to the driving force values $\tilde{F}/F_c=0.2,0.6,1$ in the upper panel and $\tilde{F}/F_c=2,6,10$ in the lower panel. The current is in units of $\tfrac{d^2 I_1(q)}{d q^2} \tfrac{dC_g}{dy} q_0\tilde{V}_g y_c$ that amounts to $\simeq 10^{-3}I_c$ 
 for the parameter set chosen.
 \label{fig:asymmetric}}
 \end{figure}

The expression for this  contribution reads
\begin{align}
I_m=\ \frac{2e}{\hbar}\frac{\partial^3E_j(q,\varphi)}{\partial q^2\partial \varphi}C_g\left(\frac{dC_g}{dy}V_{g0}{\rm Re}\left\{\tilde{V}_g\tilde{y}\right\}\right)\
\label{eq:Fano-response}
\end{align}
Interestingly, it exemplifies a Fano-type dependence on the detuning that is quite different from a Lorentzian. In the linear regime,
 \begin{align}
I_m(\nu)\sim&\ \frac{\Gamma}{2}\frac{\:\nu(\varphi)}{\nu(\varphi)^2+\Gamma^2/4},
\label{eq:Iem-phasebias}
\end{align}
so that the signal changes sign at the resonance point. Fig. \ref{fig:asymmetric} illustrates the Fano-type dependence   in the  non-linear regime.  Comparing expressions (\ref{eq:Iem-phasebias})
and (\ref{eq:Im-phasebias}) we conclude that the Fano-shaped $I_m$ dominates provided $\tilde{V}_g/V_{g0} \gg (q_0/e)^{-3/2} \simeq 10^{-3}$, this is, deep in non-linear regime.

\section{D.C. voltage bias}
\label{sec:dcvoltage}

Let us turn to d.c. voltage bias. In this case, the superconducting phase  is in first approximation a linear function of time, $\varphi=\omega_j t$, $\omega_j = 2eV/\hbar$ being the Josephson frequency which corresponds to the voltage $V$ across the junction.
In the same approximation, the current is a purely oscillatory function of time. The time-dependent current can be expanded into harmonics of the Josephson frequency,
\begin{align}
\label{eq:current-time-harmonics}
I(t) = \frac{2e}{\hbar}\frac{\partial E_j}{\partial\varphi}(q,\varphi=\omega_jt) = 
\sum_{n=1} \tilde{I}_n \sin(n \omega_j t).
\end{align}
A d.c. current emerges from the feedback on Josephson dynamics: oscillatory currents produce oscillatory corrections to the phase proportional to the impedances at frequencies $n\omega_j$. Taking this into account in the first approximation in $Z(\omega)$, we arrive at
\begin{align}
I_{{\rm d.c.}} = \sum_n |\tilde{I}_n|^2\frac{ {\rm Re} Z(n\omega_j)}{V} .
\label{eq:current-impedance}
\end{align}
The above relation holds provided $\text{Re} Z \ll V/I_c$. Non-perturbative treatment of Josephson dynamics is required otherwise.

Let us consider the mechanical effects. It is important to note that under the d.c. voltage bias the Josephson force also oscillates in time, 
\begin{align}
F_j(t) = &\ -\frac{dC_g}{dy}V_{g0}\frac{\partial E_j(q, \varphi)}{\partial q}\quad . \nonumber\\ 
= &\ -\frac{dC_g}{dy}V_{g0}\ \displaystyle\sum^{\infty}_{n=1}\frac{\partial E_{j,n}(q)}{\partial q}\cos(n \omega_j t)\quad, \label{Fj2}
\end{align}
$E_{j,n}$ being the harmonics of Josephson energy.
Therefore, the force can efficiently excite the mechanical resonator provided $n \omega_j \simeq \omega_0$. Let us first concentrate on the case where the resonance frequency is matched by the first harmonics, $\omega_j \simeq \omega_0$. The detuning is defined as
$\nu = \omega_j -\omega_0$.

To start with, let us assume that the Josephson force is sufficiently weak so that the mechanical response is linear and thus given by Eq. \eqref{ytilde}.
The direct mechanical contribution to the d.c. Josephson current is obtained by averaging Eq. \ref{eq:workflow-tilde-Im}, and reads
\begin{align}
\label{eq:mech-imy}
I_m = \frac{\partial I_1}{\partial (q/e)} \left(\frac{d C_g}{d y} \frac{V_g}{e}\right)
{\rm Im} \tilde{y}
\end{align}
This can be cast to the form similar to (\ref{eq:current-impedance}),
\begin{align}
\label{eq:current-mech-impedance}
I_{{\rm mh}} = \left|\frac{\partial I_1}{\partial (q/e)}\right|^2 
\frac{ {\rm Re} Z_m(\omega_j)}{V}
\end{align}
where the current is replaced with detecting current $\partial I/\partial (q/e)$ and the "mechanical impedance" $Z_{mh}(\omega)$ is defined as 
\begin{align}
Z_{mh}(\nu) = \frac{\omega_0}{-i\nu + \Gamma/2} Z^{(0)}_m; \label{eq:Zm}\\
Z^{(0)}_m \equiv \frac{\hbar}{e^2} \left(\frac{d C_g}{d y} \frac{V_g}{e}\right)^2 \frac{\hbar}{2 M \omega_0}. \label{eq:Z0m}
\end{align}
(Here, $\nu \equiv \omega_j -\omega_0$).

This form of the presentation of the mechanical response makes evident an analogy with Fiske steps \cite{Fiske} that are observed at voltages corresponding to resonant frequencies of an electrical impedance. This may be either an impedance of external circuit or an effective impedance that is essentially contributed Josephson inductance.   

To comprehend the scale of the response, we note first that for a sufficiently well-developed Coulomb blockade $I_1 \simeq \partial I_1/\partial (q/e)$. Therefore, to compare the current(\ref{eq:current-impedance}) and the mechanical response, we need to compare $Z_{mh}$ and a typical environmental impedance. The latter can be estimated as the impedance of free space $Z_f \simeq 10^2 {\rm Ohm}$. The typical mechanical impedance far from the resonance, $Z^{(0)}_m$, should be much smaller than that.
Indeed, if we substitute the parameter set (\ref{eq:parameter-set}) into Eq. \ref{eq:Z0m} we end up with $Z^{(0)}_m = 0.7 \cdot 10^{-2} \ll Z_f$. However, the "mechanical" impedance is enhanced by a factor of $Q$ at the resonant frequency. With this, $Z_m > Z_f$ and the current peak produced by the mechanical response should exceed the background current given and be clearly observable.

The voltage dependence of d.c. current response in linear regime is determined by $\text{Re} Z_m$ and thus takes a Lorentzian shape with the half-width 
$\delta V = V/Q$. This assumes a noiseless voltage source. It is known \cite{Likharev} that the voltage noise suppresses the coherence of Josephson generation. For white noise spectrum of intensity $S_V$, the resulting line-width reads $\delta V_n = (2e/\hbar)^2 S_V$. Comparing the two, we conclude that the mechanical response will be broadened by the noise and essentially suppressed provided $\delta V_n > \delta V$, this is, $S_V > (2e/\hbar) \omega_j/Q$. This gives a condition on the possibility of detection of the mechanical response that may be challenging to meet in practical  circumstances.  

 \begin{figure}
 \includegraphics[width=0.9\columnwidth]{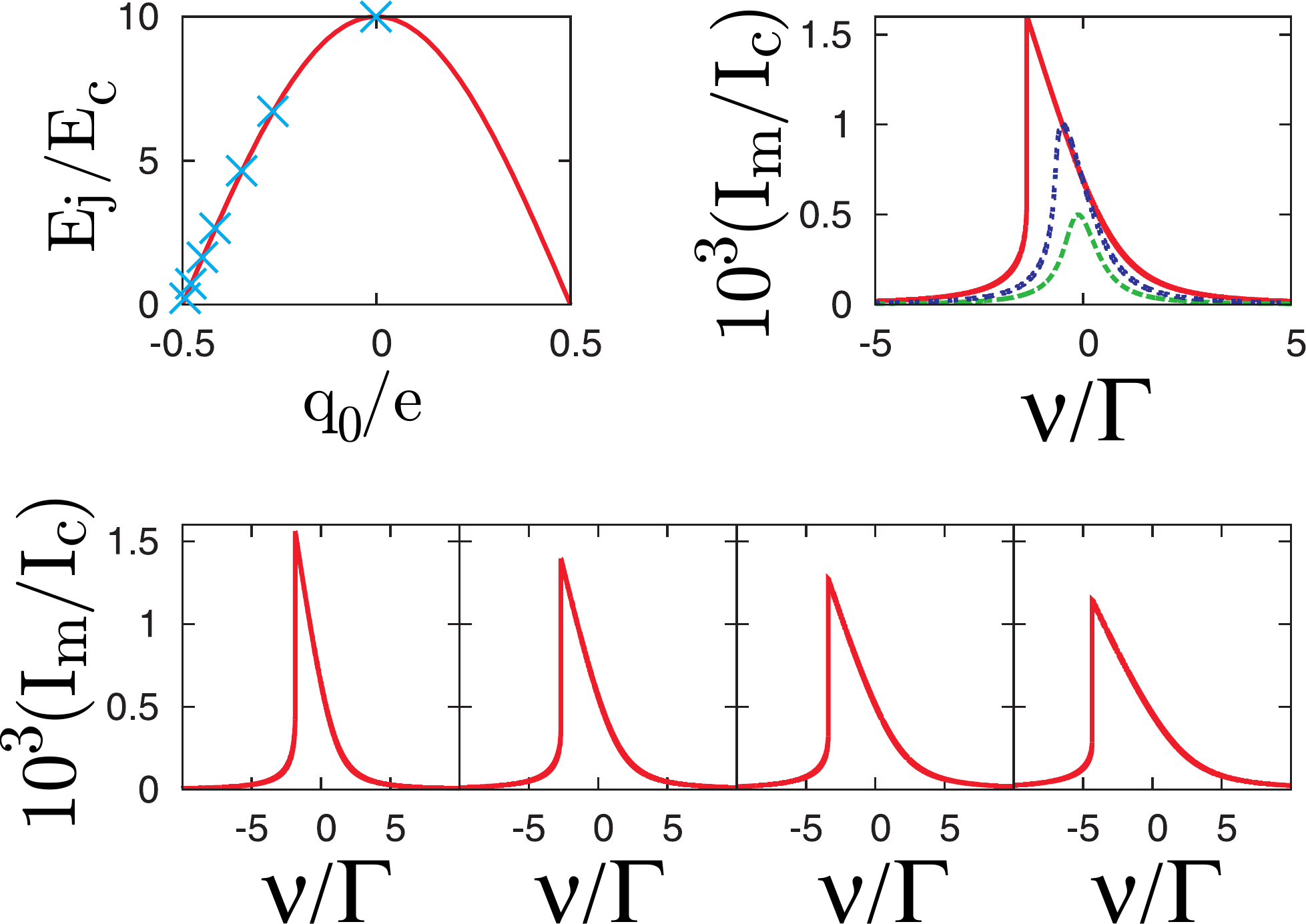}
 \caption{(Color online.)  Top left panel: The charge-dependent part of Josephson energy as  function of the gate-induced charge $q=C_g V_g$. The crosses indicate the values of $q$ that correspond to the values of Josephson force used in other panels. Top right panel: the mechanical response as function of detuning for relatively low values of the force  $F=F_j/F_c>1$, at which the response increases with increasing the force. Bottom panels: Frequency dependence at  the force values $F/F_c=3,5,7,10$ (from the left to the right panel) where the response decreases with increasing the force. 
 \label{fig:Lorentzian}}
 \end{figure}
As mentioned, the Josephson force can be big enough to exceed $F_c$, this makes it relevant to address the non-linear response as well.
We illustrate the non-linear response in Fig. \ref{fig:Lorentzian}.  To produce the Figure, we took the charge-dependent part of the Josephson energy to be of the form 
$E^{(c)}_J(q,\varphi) = E  \cos(\varphi) \cos(\pi q/e)$.  
Tuning $q$ with the d.c. gate voltage tunes the Josephson force from $0$ to some maximum value.
We chose $E$ such that the maximum force equals $10 \ F_c$ ($E = 0.23\ \mu\text{eV}$ for the parameter set in use) and compute the response using Eq. \eqref{eq:mech-imy} and Eq. \eqref{yc} at a set of the values of $q$, or, equivalently, $F_j$. The response is Lorentzian at small forces, increases and develops a jump characteristic for bistability. It is interesting to note that the response slowly decreases upon increasing $F_j$ at $F_j > 2 F_c$. This is because the response is proportional to $\text{Im} \tilde y$ that quickly decreases at big driving forces. In this limit, $I_m \propto F^{-1/3}_j$. 

\subsection{Excitation by higher harmonics}

If we take higher harmonics of current-phase characteristic into account, we note that Josephson force emerges at a set of frequencies that are integer multipliers of $\omega_j$ (Eq. \eqref{Fj2}). This implies that the resonant mechanical response can be also observed in the vicinities of  a set of discrete voltage values satisfying $\omega_j = \omega_0/n$, this is, at lower voltages than the resonance described above. The response is computed along the same lines with replacing $E_{j,1}$ by $E_{j,n}$. In linear regime,
the response reads
 \begin{align}
I_{mh} = \left|\frac{\partial I_n}{\partial (q/e)}\right|^2 
\frac{ n{\rm Re} Z_{mh}(n\omega_j)}{V}
\end{align}
(cf. Eq. \ref{eq:current-mech-impedance}, the factor $n$ in the present expression is canceled by lower voltage $V =(\hbar/2e)\omega_0/n$)
The  response scales with the relative values of the harmonics and is in principle of the same order of magnitude for several low harmonics. Its dependence on voltage in the vicinity of the resonance is similar to that discussed above and does not have to be illustrated separately.

\subsection{Parametric excitation}

For the sake of completeness, let us mention the possibility of the resonant mechanical response at {\it higher} voltages by means of parametric excitation \cite{Holmes}. Generally, parametric resonance in a non-linear oscillator is achieved by applying an a.c. driving force with frequency about a double of the resonant frequency, $\Omega \simeq 2 \omega_0$ \cite{Holmes}. In our case, this is achieved by applying a d.c. bias voltage with $\omega_j \simeq   2 \omega_0$, so that the Josephson force oscillates at $2 \omega_0$ and integer multiples of this frequency and thus provides the parametric driving required.

The response of at resonant frequency emerges provided the parametric driving force exceeds a certain threshold value, and, as in case of direct resonance, achieves values $\simeq y_c$. The point is that this threshold driving force is parametrically bigger than $F_c$, $F_t \simeq \sqrt{Q}$. For our devices at $Q =10^4$, the parametric excitation requires $E_J$ that by a factor of 30 exceed the value from the parameter set and are not practical. This is why we do not explore the regime of parametric excitation in detail. 

Besides, the manifestation of the oscillating amplitude is not as  straightforward  as in the case of direct resonance. The contribution of displacement at $\omega_0$ to the mechanical current response (\ref{eq:workflow-tilde-Im}) oscillates at the same frequency and is not readily rectified to a d.c. current. Under our assumptions, the d.c. mechanical response is dominated by the displacement oscillating at $2\omega_0$ and is by a factor of $\sqrt{Q}$ smaller than the typical responses studied in this paper.

\section{Shapiro steps at resonant driving}
\label{sec:Shapiro-res}
From now on, we turn to the situation where the setup is a.c. driven at frequency $\Omega$. As discussed in  Section 
\ref{sec:ele-setup}, in our setup this gives rise to two a.c. signals $V_g(t)=\tilde{V}_g\cos(\Omega t + \chi)$, $V_b(t)=\tilde{V}_b\cos(\Omega t)$.  The effect of $\tilde{V}_b$ is a formation of Shapiro steps \cite{Shapiro}.
 
A common approach to Shapiro steps takes into account only the first harmonics of the current-phase relation and starts with the assumption that the time-dependent superconducting phase difference can be presented as a sum of three terms
\begin{equation}
\varphi(t) =\varphi_1\sin(\Omega t) + \omega_j t + \varphi_0.\label{phiL}
\end{equation}
Here, the first term describes the a.c. driving ($\varphi_1 >0$,
$\varphi_1 = |\tilde{V}_b|/(2e/\hbar)\Omega$, the second term corresponds to a d.c. voltage $V = \omega_j/(2e/\hbar)$, and the third term is a lock-in phase important for further consideration.
With this, $\sin(\varphi)$ can be presented as a sum over harmonics 
\begin{equation}
\sin(\varphi) = \sum_{m=-\infty}^{\infty}J_m(\varphi_1)\sin(\Omega_m t+\varphi_0)\label{sinus}
\end{equation}
with $\Omega_m = m\Omega+\omega_j$. Here, $J_m$ denotes the $m$-th Bessel function of the first kind.

Shapiro steps are formed at discrete values of d.c. voltage $| \omega_j| = m \Omega$. In this case, the time-dependent current $I(t)=I_c\sin(\varphi(t))$ has a d.c. component
\begin{equation}
I_{\text{dc}} = - I_c \text{sgn}(V) J_m(\varphi_1) \sin \varphi_0
\end{equation}
Simplest assumption is ideal current bias at zero frequency and ideal voltage bias at frequencies $\simeq \Omega$. In this case, the I-V curve of a.c. driven junction consists of a series of separate pieces. At each piece (Shapiro step) the voltage is locked to one of the discrete values. The current within each piece may vary from minimum values $ I_{-}$ to the maximum value $I_+$ provided the bias current fits this interval.
In this case, the actual value of the lock-in phase $\varphi_0$ is set by the bias current.
The extremal values $I_{\pm} = \pm I_c |J_m(\varphi_1)|$ are achieved at the lock-in phases given by
\begin{equation}
\varphi^{\pm}_{0} = \mp \pi/2\ {\rm sgn}\left(VJ_m(\varphi_1)\right)\label{phpm}
\end{equation}

We in main follow this approach while admitting extreme simplifications it brings. The higher harmonics of current-phase relation and/or non-ideal voltage bias not only modify the relation between the current and lock-in phase: They also provide phase-locking at fractional ratios of $\omega_j/\Omega$ \cite{FracShapiro} and formally at all rational values of this ratio. These fractional Shapiro steps are however more sensitive to noise than the integer ones and more likely to vanish. The I-V curves of our devices do show well-developed steps at integer values of $\omega_j$ and only traces of phase-locking at intermediate values. For this reason, we do not consider fractional Shapiro steps in this paper and concentrate on integer ones where $|\omega_j| = m \Omega$.

It is advantageous to look at the mechanical response at Shapiro steps rather than at d.c. bias conditions. The external a.c. driving synchronizes Josephson oscillations. The inductive response present at Shapiro steps also reduces significantly the voltage noise at low frequencies so that it does not broaden the resonant lines. 

In this Section, we will consider the mechanical response in the simplest situation of resonant driving where the driving frequency matches the resonant frequency,
$\Omega \simeq \omega_0$.

\subsection{First step}
Let us first concentrate on the first Shapiro step, the one at voltage $2eV/\hbar =\pm \Omega$, that is the biggest in the limit of small driving voltages $\varphi_1 \ll 1$,
and determine the d.c. part of the response at the oscillating displacement $\tilde{y}$.
To represent the results, we normalize $\tilde{y}$ to the non-linearity scale $y_c$ and introduce a convenient current scale 
\begin{equation}
\bar{I} = \frac{\partial I_1}{\partial q}\frac{dC_g}{dy}V_{g0}y_c
\end{equation}

For the values of our parameter set, 
\begin{equation}
\bar{I} \simeq I_{c} (q_0/e)(y_c/L_g) = 1.1\times 10^{-3}\ I_{c}. \nonumber
\end{equation}

Making use of Eqs. \eqref{sinus} and \eqref{eq:workflow-tilde-Im}, we express the mechanical response in terms of the amplitudes $\tilde{y}$ and $\bar{I} j$ at the resonant frequency,
\begin{align}
I_{\rm mh}=&\  \frac{\bar{I}}{y_c} \text{Re}\left\{j^*\tilde{y}\right\} \label{ImAC}\\
j=&\ -i \left\{\begin{array}{cc} 
\left(J_2(\varphi_1)e^{i\varphi_0} - J_0(\varphi_1)e^{-i\varphi_0}\right) & \text{if}\ V>0\cr
\left(J_0(\varphi_1)e^{i\varphi_0} - J_2(\varphi_1)e^{-i\varphi_0}\right) & \text{if}\ V<0
\end{array}\right.
\end{align}

This displacement is a response on the force at resonant frequency which is a sum of Josephson force and gate force. 
The time-dependent Josephson force is expanded in harmonics in the form
\begin{align}
F_j(\varphi) =&\ \bar{F}_j\cos(\varphi(t)),\label{JosephsonF}\\
 = &\ \bar{F}_j\sum_{m=-\infty}^{\infty}J_m(\varphi_1)\cos(\Omega_m t+\varphi_0)\nonumber\\
\bar{F}_j=&\  -\frac{dC_g}{dy}V_{g0}\frac{\partial E_{1,j}(q)}{\partial q} \simeq \frac{q_0}{e}\frac{E_j}{L_g}.\nonumber
\end{align}
Its amplitude at resonant frequency is contributed by the terms $m=0,2$ 
and reads 
\begin{align}
\tilde{F}_j =&\ \bar{F}f; \\
f =&\ \left\{\begin{array}{cc} 
\left(J_2(\varphi_1)e^{i\varphi_0} + J_0(\varphi_1)e^{-i\varphi_0}\right) & \text{if}\  V>0\cr
\left(J_0(\varphi_1)e^{i\varphi_0} + J_2(\varphi_1)e^{-i\varphi_0}\right) & \text{if}\ V<0
\end{array}\right.
\end{align}

Let us discuss first the relative scale of the gate force in comparison with the Josephson force. 
It may seem there is none, and varying the a.c. gate voltage $\tilde{V}_g$ one can achieve any ratio $\simeq \tilde{V}_g/E_J$ between the forces. However, we should take into account the fact that in our setups a.c. driving also induces an appreciable bias voltage $\tilde{V}_b$. If the oscillating phase $\varphi_1$ produced by this voltage becomes large Shapiro steps can hardly be observed. It is in general reasonable to expect $\tilde{V}_g \simeq \tilde{V}_{b}$.
In this case, $\varphi_1 \simeq 1$ corresponds to $F_g/F_j \simeq \hbar \omega_0/E_j$. The latter ratio is typically $10^{-2}$ in our setups (for our parameter set it is $\hbar \omega_0/E_j=2.7\times 10^{-2}$). This implies that typically we can disregard a.c. gate force in comparison to the Josephson force. We will analyze this case first and consider the effect of the gate force in the end of the Subsection.

With this, the mechanical response is given by
\begin{align}
I_{{\rm mh}} = \frac{\bar{I} \bar{F}}{F_c} \text{Re}\{j^*f R\} = \notag \\
\frac{\bar{I} \bar{F}}{F_c} \left( \text{sgn}{V} \left(J^2_0(\varphi_1)-J^2_2(\varphi_1) \right)\text{Im}(R) \right. \notag \\
-\left. J_0(\varphi_1)J_2(\varphi_1) \sin(2\varphi_0) \text{Re}\{R\}\right) \label{eq:firststep-Jo}
\end{align}
Here, $R \equiv R(\nu/(\Gamma/2), (F_j/F_c) |f|)$ defined by Eq. \eqref{yc} gives the non-linear mechanical response. The expression is naturally separated onto two terms. The first term is proportional to $\text{Im}(R)$ and therefore exhibits a Lorentz-like dependence on frequency.
It does not depend on the lock-in phase and can be regarded as a {\it shift} in the current. 
Owing to the shift, the maximum and minimum currents $I^{\pm}$ at the step are no more opposite: the mechanical effect breaks the symmetry of the Shapiro step. The shift is however opposite for opposite voltages. The origin of the shift may be traced to the Fiske response (Eq. \ref{eq:current-mech-impedance}) formed at $\omega_j \simeq \omega_0$ in the absence of the a.c. driving. Indeed, in the limit of vanishing $\varphi_1$ the first term in the mechanical response does not vanish: rather, it approaches the expression (\ref{eq:current-mech-impedance}). So it looks like the Fiske response persists also for well-developed Shapiro steps and contributes to the current at the step. This suggest perhaps the easiest way to observe and identify the mechanical response: measure maximal and minimal currents at a step as function of the a.c. frequency. In the rest of the paper we thus mainly concentrate on the modification of extremum currents.

The second term in Eq. \eqref{eq:firststep-Jo} cannot however be observed in this way.
In ideal current bias conditions at low frequency,  the second term in the current response fact amounts to a shift of the lock-in phase. Indeed, since the current at the step as function of $\varphi_0$ reads as $I(\varphi) = - \text{sgn}(V) J_1(\varphi_1) $,  the second term can be seen as a modification of the lock-in phase at constant bias current which does not depend on this current,
\begin{equation}
\left(\Delta \varphi_0\right)_{\text{mh}} =  -\text{sgn}(V)\frac{\bar{I} \bar{F}}{ I_c F_c}  \frac{J_0(\varphi_1)J_2(\varphi_1)}{J_1(\varphi_1)}. \text{Re}(R)
\end{equation}
This response is of Fano-type. Since such shift of the phase does not modify the values of the current extrema, the effect cannot be observed in the course of two-terminal electrical measurement in our setup. The shift of the lock-in phase can be however revealed if the Josephson junction under consideration is a part of a SQUID, or with the aid of lock-in measurement at non-resonant a.c. frequency. 

With respect to this, we ought to mention yet another effect of Josephson force manifesting itself in the mechanical response considered. In fact, the situation at a Shapiro step is similar to the phase bias conditions considered in Section \ref{sec:phasebias}, with lock-in phase playing the role of $\varphi$. We thus expect $\varphi_0$-dependent shift of the resonance frequency. The static Josephson force at a Shapiro step is given by
\begin{equation}
F_j = \bar{F}\ \text{sgn}(V) J_1(\varphi) \cos(\varphi_0)
\end{equation}
The frequency shift caused by this force thus reads
\begin{equation}
\Delta \omega_0(\varphi_0) = \text{sgn}(V) (\Delta \omega_0)_\text{max} J_1(\varphi) \cos(\varphi_0).
\end{equation}
Here $(\Delta \omega_0)_\text{max}$ is a maximum frequency shift in the absence of the a.c. driving, given by Eq. \eqref{shift}.
The frequency shift vanishes at extremum points $\varphi_0^{\pm}$ and therefore cannot be observed by measuring the extrema of the current.

 \begin{figure}
 \includegraphics[width=0.8\columnwidth]{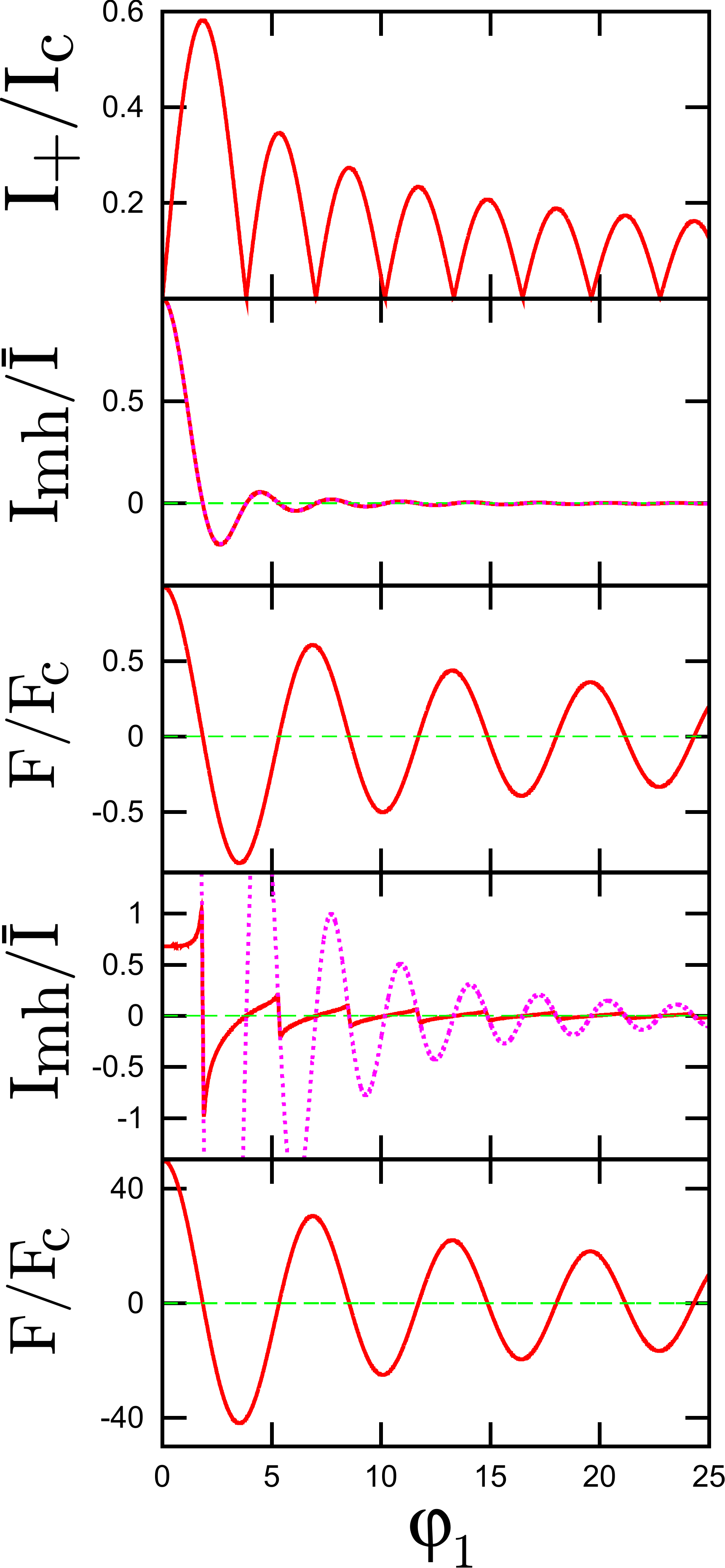}
 \caption{(Color online.) The mechanical response at resonant driving $\Omega \simeq \omega_0$ and at the first Shapiro step $V_0=2e\Omega/\hbar$ versus the oscillating phase $\varphi_1$. The first (upper) plot  gives the maximum current at the step. 
 The second and fourth plots give the mechanical response defined as the extremum of the modification of this maximum current over the frequencies near the resonance, at maximum Josephson forces $\bar{F}=F_c$ and $\bar{F}=50 F_c$, respectively. The actual amplitudes of the resonant Josephson forces are given at the lower plots, respectively third and fifth. }
 \label{fig:firststep-FJ}
 \end{figure}

\begin{figure}
 \includegraphics[width=0.7\columnwidth]{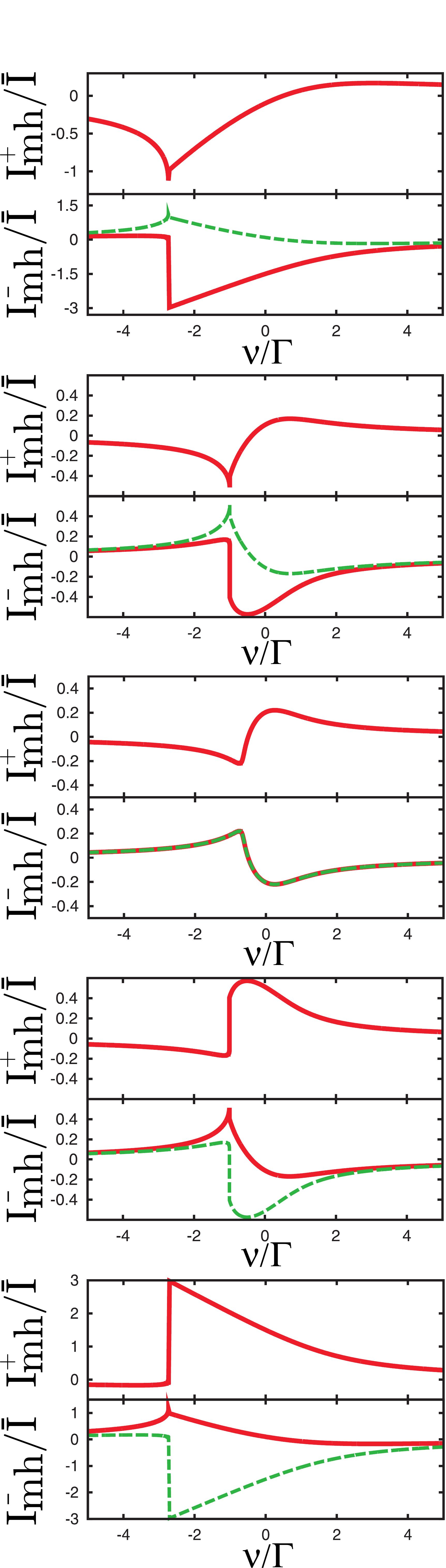}
 \caption{(Color online.) The effect of gate force. For all plots, the gate force is fixed to $F_g=F_c$. In the plots from top to bottom the maximum Josephson force  $\bar{F}$ assumes the values $\bar{F}/F_c=-5,-1,0,1,5$. We choose $\chi=0$ and $\varphi_1=1$. Dashed lines in the plots for $I^{-}$ give values opposite to the corresponding $I^{+}$, to stress the symmetry or asymmetry of the response. }
 \label{fig:Shapirowithgate}
 \end{figure}

 We illustrate the mechanical response in Fig. \ref{fig:firststep-FJ}. In this Figure as well in all subsequent Figures except Fig. \ref{fig:Shapirowithgate}, we concentrate on the modification of the maximum current on the step. Instead of presenting the (rather trivial
Lorentz-like) frequency dependence of this modification, we give the extremum of this modification over the frequency range and plot it versus $\varphi_1$. The extremum is proportional to the maximum of $\text{Im}(R)$ over the frequency. We shall note that the dependence of this maximum on the force is rather specific one: it is a constant until the bistability threshold $F=1.24 F_c$, has a cusp at this value of force, and decreases monotonously at higher forces. This accounts for rather strange appearance of the response curves. If the Josephson force is smaller than the bistability threshold, they coincide with the linear response given by the dotted curves. Otherwise, the response is smaller than linear one and exhibits kinks.

The so-defined maximum response is plotted in Fig. \ref{fig:firststep-FJ} for two values of $\bar{F}$, those correspond to slightly and strongly non-linear regime, respectively. In both cases, the response vanishes when the width of Shapiro step reaches maximum, or becomes zero (except $\varphi_1=0$). In slightly non-linear regime, the response reaches maximum value at $\varphi_0$. Upon increasing $\varphi_1$, it exhibits  Bessel-like oscillations. The envelop of these oscillations shrinks with increasing $\varphi_1$. This shrinking is much faster than that for either step width or Josephson force. In strongly non-linear regime, the amplitude of the response is determined by competition of two factors: it is increased by the bigger value of the detecting current, and decreased owing to smaller imaginary part of oscillating displacement at higher Josephson forces.
   
Let us turn to the effect of the gate force. The Josephson force of the kind considered can change sign and therefore be tuned to zero by tuning $q$. In the vicinity of this particular $q$, the gate force should compete with the Josephson one and eventually dominate. The full amplitude of the force at the resonant frequency then
reads 
\begin{equation}
\tilde{F} = \bar{F} f + F_g \exp(-i\chi),
\end{equation}
the frequency shift $\chi$ between the bias and the gate voltage being a relevant parameter.

The mechanical response is given by Eq. \eqref{eq:firststep-Jo} where $R$ depends on the full force plus 
an addition proportional to $F_g$,
\begin{align}
I^{(g)}_{{\rm mh}} = \frac{\bar{I} F_g}{F_c}  \text{Re}\{j^*\exp(-i\chi) R\} = \notag \\
\frac{\bar{I} \bar{F}}{F_c} \left( -(J_0(\varphi_1)\sin(\varphi_0+\chi) +J_2(\varphi_1)\sin(\varphi_0-\chi) \right)\text{Re}(R) \notag \\
\left.(J_0(\varphi_1) \cos(\varphi_0+\chi) -J_2(\varphi_1)\cos(\varphi_0-\chi)) \text{Im}(R)\right) 
\label{eq:firststep-gate}
\end{align}
The last equation holds for $V>0$. The expression for $V<0$ is obtained by interchanging $J_0$ and $J_2$. 
Evaluating this at the extremum points of lock-in phase, $\varphi_0^{\pm}$, we obtain
\begin{align}
I^{(g)\pm}_{{\rm mh}} = &\ \mp \text{sgn}(VJ_1(\varphi))\frac{\bar{I} F_g}{F_c} \\
\ &\ (J_0(\varphi_1) +J_2(\varphi_1)) \text{Re}\{R\exp(-i\chi)\}\nonumber
\end{align}
Therefore, the contribution of the gate force to extremum currents is not like a shifts: rather, it modifies the width of the step. These terms are even in voltage and display a mixture of Fano-type and Lorentzian-type  response as function of frequency, this being tuned by the phase $\chi$.

 To illustrate a rather complex interplay of Josephson and we plot in Fig. \eqref{fig:Shapirowithgate} the frequency dependence of the mechanical response for a constant $V_g$ and a set of values of $F_j$ that pass zero.  The plots show the modifications of extremum currents $I^{\pm}$. These modifications are the same for the Josephson force contribution and opposite for the gate force contribution. Besides, the frequency dependence is Fano-like for the gate force contribution and Lorenz-like for the Josephson force contribution. In the central plot, the  Josephson force contribution is absent, the modifications of $I^{\pm}$ are opposite, and the frequency dependence is Fano-like. Upon increasing the Josephson force, these features are transformed into the opposite ones. The plots are symmetric upon simultaneous change of signs of the current and the Josephson force.
  
\subsection{Higher steps}

At the same conditions of the resonant driving, we analyze the mechanical response at other Shapiro steps $m>1$, those correspond to higher voltages $|V|= m \hbar \omega_0 /2e$.  Both the amplitudes of the detecting current and the Josephson force display a complex dependence on the step number $m$ and the oscillating phase $\varphi_1$. They are given by
\begin{align}
j = -i \left(J_{-1+\bar{m}} e^{i\varphi_0} - J_{1+\bar{m}} e^{-i\varphi_0} \right)\notag \\
f= J_{-1+\bar{m}} e^{i\varphi_0} + J_{1+\bar{m}} e^{-i\varphi_0}\notag 
\end{align} 
where the dependence on the sign of the voltage is incorporated into $\bar{m} \equiv -\text{sgn}(V)m$.
We consider only the situation when Josephson force dominates. With this, we obtain a relation similar to Eq. \eqref{eq:firststep-Jo}:
\begin{align}
I_{{\rm mh}} = \frac{\bar{I} \bar{F}}{F_c} \text{Re}(j^*f R) = \notag \\
\frac{\bar{I} \bar{F}}{F_c} \left( \text{sgn}{V} \left(J^2_{m-1}(\varphi_1)-J^2_{m+1} (\varphi_1) \right)\text{Im}(R) \right. \notag \\
-\left. J_{m-1}(\varphi_1)J_{m+1}(\varphi_1) \sin(2\varphi_0) \text{Re}(R)\right) \label{eq:allsteps-Jo}
\end{align}
As in the first step, the response consists of two terms. The first one gives a shift in the current, and gives a modification of the maximum and minimum currents at the step, this is to be measured. As in the previous case, the shift is odd in voltage. However, its $\varphi_1$-dependence is quite rather distinct.

The measuring of the mechanical response at higher steps is important to check the consistency of results and thereby unambiguously identify the mechanism of the response. The characteristic dependences on $\varphi_1$ make the identification easy.

\begin{figure}
 \includegraphics[width=0.9\columnwidth]{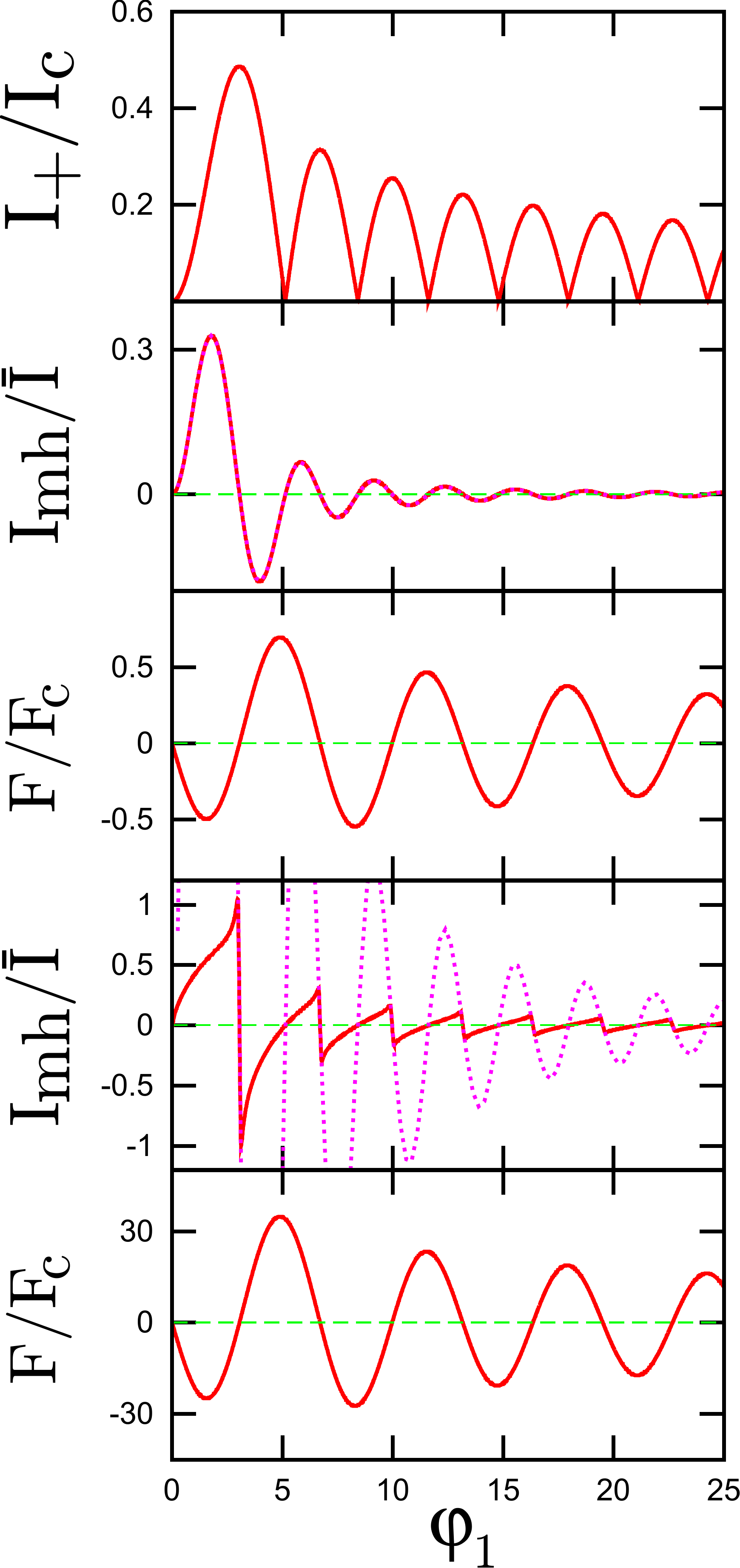}
 \caption{(Color online.) The maximum current at the Shapiro step, the mechanical response and  the force versus $\varphi_1$ for $\bar{F} = F_c$ and  $\bar{F} = 50 F_c$ at the second Shapiro step $V_0=4e\Omega/\hbar$.}
 \label{fig:higher-steps-second}
 \end{figure}

\begin{figure}
 \includegraphics[width=0.9\columnwidth]{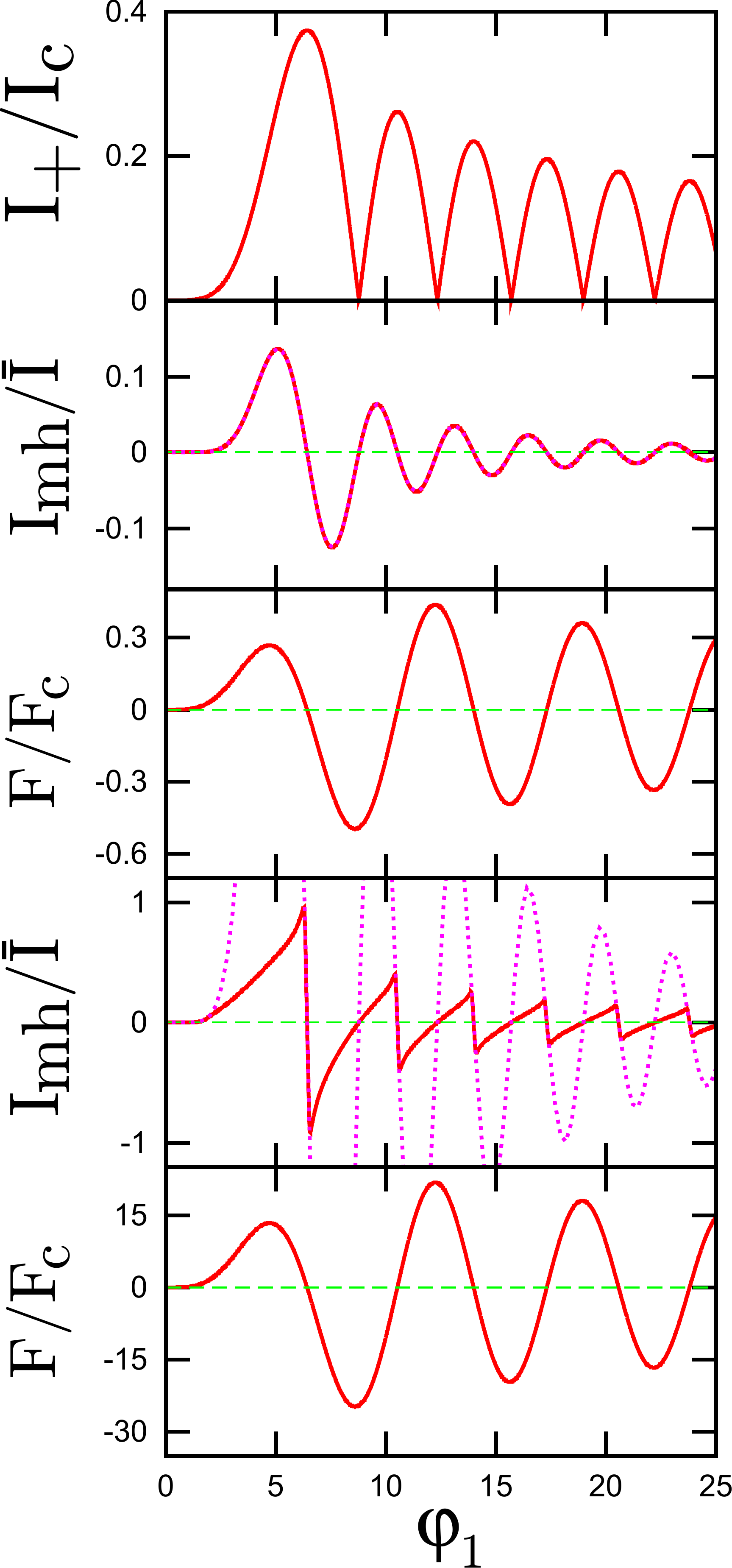}
 \label{fig:higher-steps-fifth}
 \caption{(Color online.) The same as in Fig. \ref{fig:higher-steps-second} at the fifth Shapiro step $V_0=10e\Omega/\hbar$.}
 \end{figure}

We illustrate the response for higher steps in Fig.  \ref{fig:higher-steps-second} (second step) and Fig. \ref{fig:higher-steps-fifth} (fifth step). In both cases, the response correlates with the Shapiro step width given in the upper plots: it vanishes when the width achieves a maximum or becomes zero. In distinction from the first step, the responses vanish at vanishing $\varphi_1$. Their typical values are of the same order. However, the envelopes of the responses decrease rather slow with increasing $\varphi_1$.

\section{Shapiro steps at non-resonant driving}
\label{sec:Shapiro-non-res}
In the previous Section, we concentrate on the case when the driving frequency $\Omega$ matches the resonant frequency of the mechanical oscillator. It is not a necessary condition for an efficient excitation of the resonant mode.
The Josephson dynamics at Shapiro steps are essentially non-linear. As a consequence, the spectrum of current oscillations contain all higher harmonics $n\Omega$ of the driving frequency $\Omega$. The same pertains the Josephson force. Therefore, the resonator can be efficiently excited for $\Omega=\omega_0/N$, $N >1$ being an integer number. At any given $N$, the resonant conditions are achieved for any Shapiro step number $m$, and  thus for voltages $2eV/\hbar =\omega_j= (m/N) \omega_0$.

 \begin{figure*}
 \includegraphics[width=0.9\textwidth]{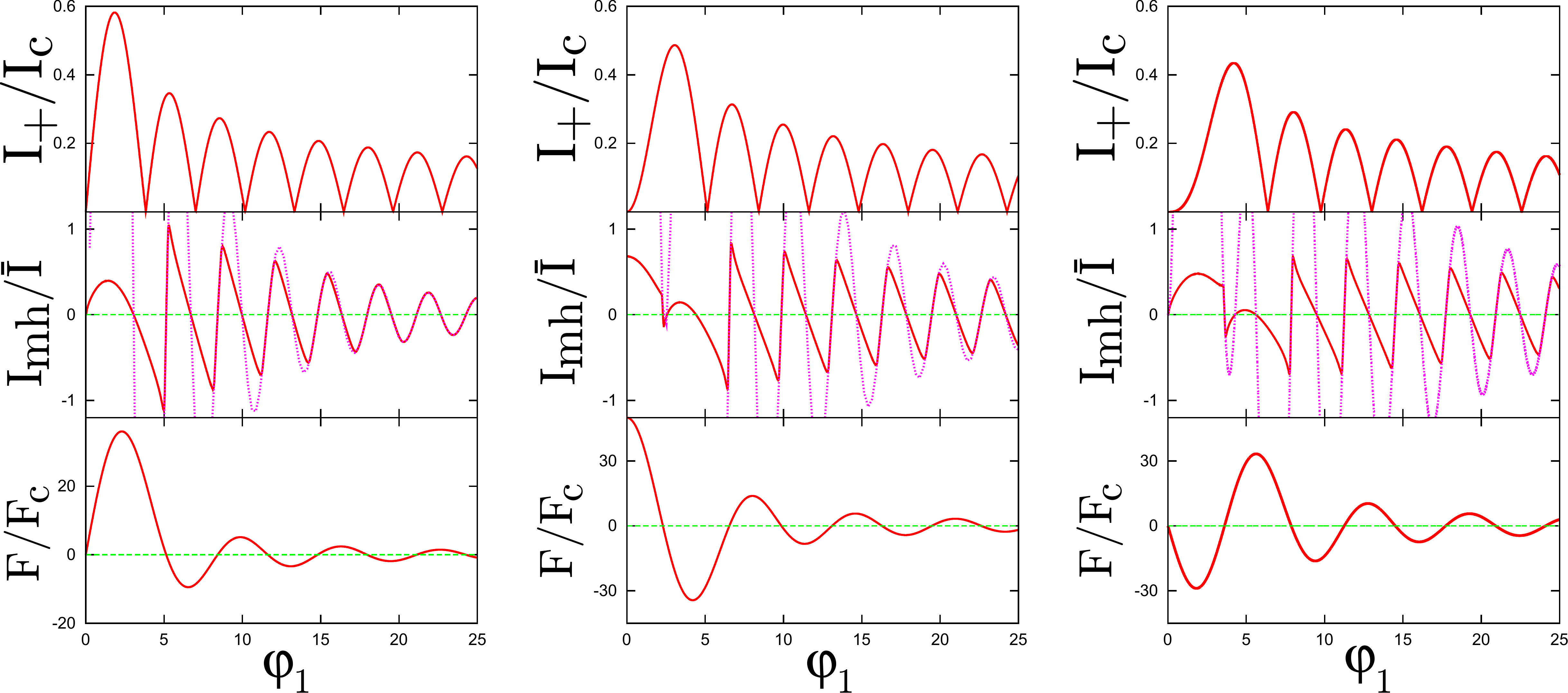}
 \caption{(Color online.) The mechanical response at the non-resonant driving. Here, the a.c. driving frequency is $\Omega \simeq \omega_0/2$, corresponding to $N=2$. From left to right the three columns correspond to Shapiro steps $m=1,2,3$. Plotted are the maximum current at the Shapiro step, the mechanical response and the amplitude of the force. The maximum of the Josephson force was set to $\bar{F}=50F_c$ for all plots.}
 \label{fig:OffResonantResponse1}
 \end{figure*}
 \begin{figure*}
 \includegraphics[width=0.9\textwidth]{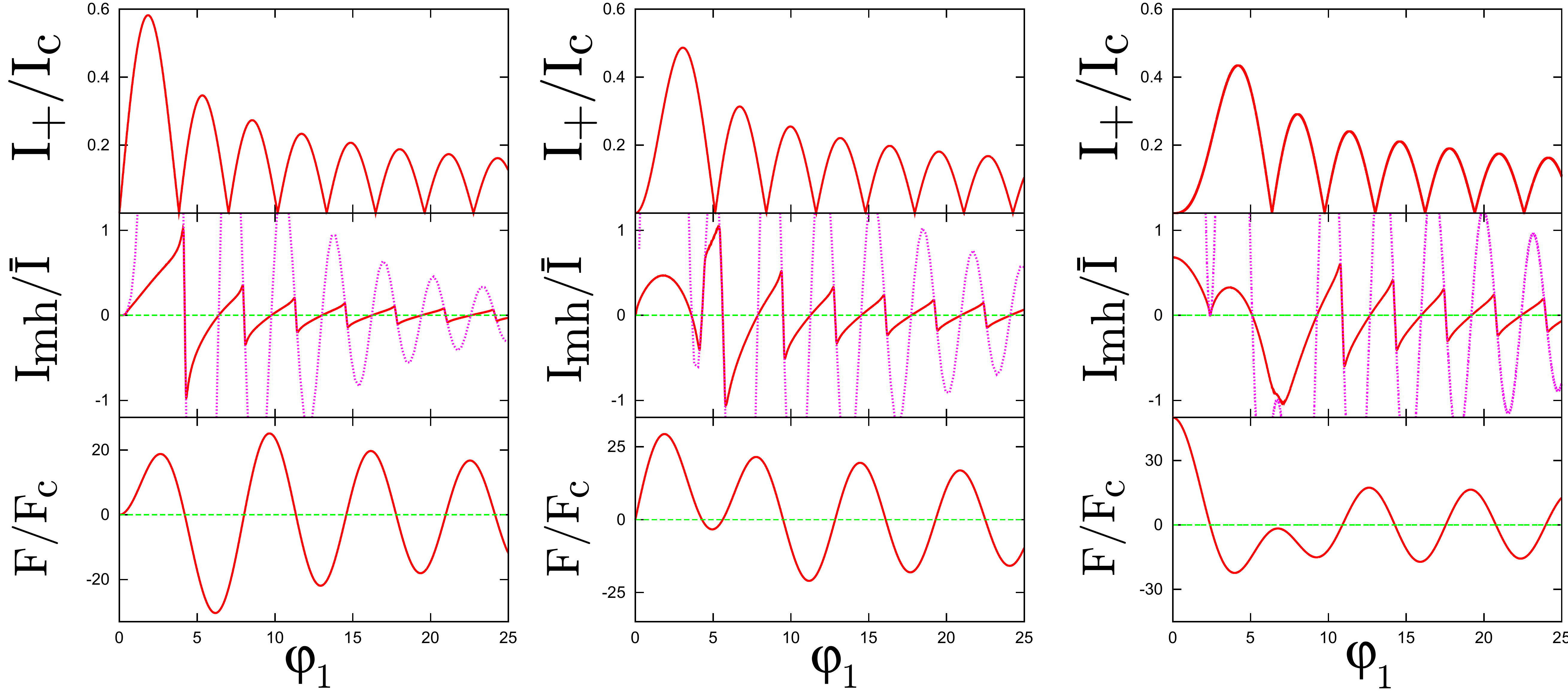}
 \caption{(Color online.) The mechanical response at the non-resonant driving for $N=3$. Except this, all other parameters are the same as in Fig. \ref{fig:OffResonantResponse1}}
 \label{fig:OffResonantResponse2}
 \end{figure*}

These non-resonant driving conditions are advantageous
for observation of the Josephson force
since the a.c. gate voltage force  is not in the resonance, does not cause any appreciable displacement and therefore  
does not mask the effect of the Josephson force. In this short Section, we will thus concentrate on the case of the non-resonant driving $\Omega = \omega_0/N$.

The amplitudes of the detecting current and the Josephson force  depended not only on  the step number $m$ and the oscillating phase $\varphi_1$, but also on $N$. They are given by
\begin{align}
j = -i \left(J_{-N+\bar{m}} e^{i\varphi_0} - J_{N+\bar{m}} e^{-i\varphi_0} \right)\notag \\
f= J_{-N+\bar{m}} e^{i\varphi_0} + J_{N+\bar{m}} e^{-i\varphi_0}\notag 
\end{align} 
where the dependence on the sign of the voltage is again incorporated into $\bar{m} \equiv -\text{sgn}(V)m$.

Since the gate force is absent, the response is given by a relation similar to Eq. \ref{eq:firststep-Jo} that contains the Josephson force only:
\begin{align}
I_{{\rm mh}} = \frac{\bar{I} \bar{F}}{F_c} \text{Re}(j^*f R) = \notag \\
\frac{\bar{I} \bar{F}}{F_c} \left( \text{sgn}{V} \left(J^2_{m-N}(\varphi_1)-J^2_{m+N} (\varphi_1) \right)\text{Im}(R) \right. \notag \\
-\left. J_{m-N}(\varphi_1)J_{m+N}(\varphi_1) \sin(2\varphi_0) \text{Re}(R)\right). \label{eq:Nallsteps-Jo}
\end{align}
It is again separated onto two terms discussed above, only the first term being responsible for the modification of the extremum currents of the Shapiro steps.

We illustrate the dependences on $\varphi_1$ in Fig. \ref{fig:OffResonantResponse1} (for $N=2$) and Fig. \ref{fig:OffResonantResponse2} (for $N=3$) for the first three steps with $m=1,2,3$. The vertical arrangement of the plots is the same as in the previous Figures except we chose a single value of the maximum Josephson force $\bar{F} = 50 F_c$ that brings us deep into the non-linear regime. In this regime, the response is of the same order of magnitude for all steps and ratios $N$, while retaining unique $m,N$ specific dependence on $\varphi$ that can be used for identification of the effect and the characterization of the Josephson force.

\section{Conclusions}
\label{concl}

In conclusion, we have studied Josephson junction dynamics affected by
excitation of a mechanical resonator.
We have demonstrated that the mechanical oscillations
can be rectified giving rise to an additional d.c. current 
that can be used for detection. The mechanical response is
proportional to the oscillation amplitude, and is estimated 
as $\bar{I}_\text{mh} \simeq I_c (q_0/e) (y/L)\simeq 10^{-3}I_c(y/y_c)$. 
The resonator can be driven by the a.c. voltage 
applied to the gate electrode as well as an additional mechanical
force, termed
the Josephson force, that depends on the superconducting phase difference at 
the junction. We estimate the Josephson force as $\bar{F}_j\simeq (q_0/e)(E_j/L_g)$ and show that it is sufficiently strong to drive the mechanical resonator into the non-linear regime. We also show that it is typically larger than
magneto-induced force proposed in \cite{Sonne}.

We have presented a general and detailed analysis of the coupling between electrical and mechanical degrees of freedom, discussing the competing non-linearities. This analysis is applied to a Josephson device with a suspended CNT resonator, where we show that the intrinsic non-linearity scales dominate those arising from the coupling.

We have provided analytical formulas for the response of the device to mechanical excitations in a wide interval of the excitation strengths and for various biasing schemes. We discuss distinct frequency dependencies, Lorentz and Fano-like, of the mechanical response both for linear and non-linear regimes  and show how these arise based on the nature of the resonant mechanical force. In the case of a phase biased junction we show that the resonant frequency of the mechanical mode acquires a measurable phase-dependent shift (see Fig. \ref{fig:shift}). 

We have discussed conditions of detecting the enhanced mechanical response arising when the Josephson frequency matches the resonance frequency of the mechanical mode. We reasoned that the regime of Shapiro steps is advantageous, since the fluctuations of the voltage drop over the junction are suppressed. We provided expressions for the mechanical response in the regime of Shapiro steps and demonstrated that it manifests as modifications of the extrema of the steps. We show that the mechanical mode can be efficiently excited not only by resonant a.c. signals, but also by a.c. signals with frequencies close to an integer fraction of the mechanical resonance frequency. Our preliminary experimental results confirm this behavior; these will be reported elsewhere.

\begin{acknowledgments}
The authors would like to thank G. A. Steele, S. M. Frolov, L. P. Kouwenhoven, and H. S. J. van der Zant for many useful discussions. This research was supported by the Dutch Science Foundation NWO/FOM.
\end{acknowledgments}

\end{document}